\documentclass[reqno,11pt]{amsart}
\usepackage[utf8]{inputenc}
\usepackage{enumitem}
\usepackage{graphicx}
\usepackage{amscd}
\usepackage{slashed}
\usepackage{amssymb}
\usepackage{tikz}
\usepackage{tikz-cd}
\usepackage{mathtools} 
\usepackage[mathscr]{eucal}
\numberwithin{equation}{section}
\allowdisplaybreaks[1]

\title[Gauge Fixing for Causal Fermion Systems]{A Gauge Fixing Procedure for \\ Causal Fermion Systems}

\author[F.\ Finster]{Felix Finster}

\author[S.\ Kindermann]{Sebastian Kindermann \\ \\ August 2019 / May 2020}
\address{Fakult\"at f\"ur Mathematik \\ Universit\"at Regensburg \\ D-93040 Regensburg \\ Germany}
\email{finster@ur.de, Sebastian.Kindermann@web.de}

\newtheorem{Def}{Definition}[section]
\newtheorem{Thm}[Def]{Theorem}
\newtheorem{Prp}[Def]{Proposition}
\newtheorem{Lemma}[Def]{Lemma}

\newcommand{\Thanks}{\vspace*{.5em} \noindent \thanks}
\newcommand{\beq}{\begin{equation}}
\newcommand{\eeq}{\end{equation}}
\newcommand{\Proof}{\begin{proof}}
\newcommand{\QED}{\end{proof} \noindent}

\newcommand{\la}{\langle}
\newcommand{\ra}{\rangle}

\newcommand{\Sl}{\mathopen{\prec}}
\newcommand{\Sr}{\mathclose{\succ}}
\newcommand{\C}{\mathbb{C}}
\newcommand{\R}{\mathbb{R}}
\newcommand{\1}{\mbox{\rm 1 \hspace{-1.05 em} 1}}
\newcommand{\Z}{\mathbb{Z}}
\newcommand{\N}{\mathbb{N}}
\renewcommand{\H}{\mathscr{H}}
\renewcommand{\O}{\mathscr{O}}

\newcommand{\F}{{\mathscr{F}}}

\renewcommand{\O}{{\mathscr{O}}}

\newcommand{\Lin}{\text{\rm{L}}}
\newcommand{\U}{\text{\rm{U}}}

\newcommand{\scrM}{\myscr M}

\newcommand{\Pdd}{\mbox{$\partial$ \hspace{-1.2 em} $/$}}
\newcommand{\bitem}{\begin{itemize}[leftmargin=2em]}
\newcommand{\eitem}{\end{itemize}}

\DeclareFontFamily{OT1}{rsfso}{}
\DeclareFontShape{OT1}{rsfso}{m}{n}{ <-7> rsfso5 <7-10> rsfso7 <10-> rsfso10}{}
\DeclareMathAlphabet{\myscr}{OT1}{rsfso}{m}{n}

\DeclareMathOperator{\tr}{tr}
\DeclareMathOperator{\Symm}{\mbox{\rm{Symm}}}

\newcommand{\p}{\mathfrak{p}}
\newcommand{\q}{\mathfrak{q}}

\begin{document}

\maketitle

\begin{abstract}
Causal fermion systems incorporate local gauge symmetry in the sense that
the Lagrangian and all inherent structures are invariant under local phase transformations
of the physical wave functions.
In the present paper it is explained and worked out in detail that, despite this local gauge freedom,
the structures of a causal fermion system give rise to
distinguished gauges where the local gauge freedom is fixed completely up to global
gauge transformations.
The main method is to use spectral and polar decompositions of operators on Hilbert spaces and
on indefinite inner product spaces.
We also introduce and make use of a Riemannian metric which is induced on the manifold of all
regular correlation operators by the Hilbert-Schmidt scalar product.
Gaussian coordinate systems corresponding to this Riemannian metric are constructed.
Moreover, we work with so-called wave charts where the physical wave functions
are used as coordinates. Our constructions and results are illustrated in the example
of Dirac sea configurations in finite and infinite spatial volume.
\end{abstract}
\tableofcontents

\newpage

\section{Introduction}
The {\em{local gauge freedom}} of electrodynamics is based on the observation that transforming
the electromagnetic potential~$A$ in Minkowski space by the derivative of a real-valued function~$\Lambda$,
\beq \label{gauge1}
A_j(x) \rightarrow A_j(x) + \partial_j \Lambda(x) \:,
\eeq
does not change the electromagnetic field tensor and thus has no effect on any observable quantities
(see for example~\cite[Section~I.2]{landau2}).
In quantum mechanics, the gauge transformation~\eqref{gauge1} must be complemented by
a local phase transformation of the wave function~$\psi$ (see~\cite[Section~XV.111]{landau3},
\cite[Section~4.1]{peskin+schroeder} or~\cite[Section~2.6]{sakurai}),
\beq \label{gauge2}
\psi(x) \rightarrow e^{i \Lambda(x)}\: \psi(x) \:.
\eeq
The connection between these two transformation laws can be understood most easily if the
electromagnetic potential is combined with the partial derivatives to gauge-covariant derivatives~$D_j$ by
\beq \label{Djtrans}
D_j := \partial_j - i A_j \:,
\eeq
because then~\eqref{gauge1} and~\eqref{gauge2} give rise to the simple transformation law
\[ D_j \psi(x) \rightarrow e^{i \Lambda(x)}\: D_j \psi(x) \:. \]
The local gauge principle states that local gauge transformations lead to equivalent
formulations of the physical system. In its generalization to non-Abelian gauge theories,
the local gauge principle is one of the cornerstones of modern physics.

In most applications and calculations, the local gauge freedom is inconvenient because
of the resulting non-uniqueness of the gauge potential and the gauge phases.
Therefore, it is often desirable to {\em{fix the gauge}}, for example by choosing the
Lorenz, Coulomb or radiation gauges. The general strategy of
a gauge-fixing procedure is to use the local gauge freedom in order to arrange that the
gauge potential has a particularly simple or convenient form.
Many gauge-fixing procedures do not fix the gauge completely, but only {\em{partially}} up to a remaining
residual gauge freedom. In particular, the residual gauge freedom typically includes the
{\em{global}} gauge transformations (i.e.\ transformations of the form~\eqref{gauge1} and~\eqref{gauge2}
with~$\Lambda$ a constant).

{\em{Causal fermion systems}} are a recent approach to fundamental physics
(see the basics in Section~\ref{secprelim}, the reviews~\cite{dice2014, nrstg, review}, the textbook~\cite{cfs}
or the website~\cite{cfsweblink}). It is a major feature of the approach that the physical system
is encoded in a measure~$\rho$ on a set of bounded linear operators~$\F \subset \Lin(\H)$
of a Hilbert space~$(\H, \la .|. \ra_\H)$ (for details see the abstract definition in Section~\ref{seccfs}).
The causal fermion system~$(\H, \F, \rho)$ is defined in a manifestly gauge-invariant manner
(see also Section~\ref{secquantumcfs}).
Nevertheless, when representing the vectors in~$\H$ as wave functions in spacetime,
local gauge freedom arises as the freedom in choosing basis representations of
the spinors at each spacetime point (for details see Section~\ref{secgaugecfs}).
This raises the question if and to what extent
the structures of a causal fermion system make it possible to fix the gauge.
In preparation for tackling this question, in~\cite[Section~6]{perturb} it was noted that in the setting of
causal fermion systems, gauge freedom corresponds to the freedom in choosing charts on~$\F$.
Based on this observation, in~\cite[Section~5.3]{nrstg}
a gauge-fixing procedure was proposed for causal fermion systems.
In the present paper we work out this procedure in detail and clarify how it is related
to local gauge freedom and gauge fixing in electrodynamics.
For clarity, we point out that we always restrict attention to the case that the Hilbert space~$\H$
is {\em{finite-dimensional}} (for the infinite-dimensional case see~\cite{banach}).
In order to clarify the above-mentioned connection between gauges and charts on~$\F$,
we work with so-called {\em{wave charts}}, where the spinorial wave functions are used
as the coordinates. We obtain the following two main results:
\bitem
\item[(i)] In a neighborhood of any spacetime point there is a canonical distinguished gauge
which is unique up to global gauge transformations.
\item[(ii)] Treating the electromagnetic field perturbatively, there is a canonical way
to fix the local gauge freedom to every order in perturbation theory, again up to global gauge transformations.
\eitem
These distinguished gauges are described mathematically by so-called {\em{symmetric
wave charts}} (see Theorem~\ref{thmchart1}). The constructions and results are illustrated 
for Dirac systems in Minkowski space in Section~\ref{secfixst} (result~(i)) and Section~\ref{secfixpe} (result~(ii)).

The paper is organized as follows. Section~\ref{secprelim} gives a brief but self-contained
introduction to causal fermion systems and provides the necessary background material.
In Section~\ref{secmanifold} it is shown that the subset~$\F^{p,q}$
of symmetric linear operators of rank at most~$p+q$ which (counting multiplicities) have~$p$ positive
and~$q$ negative eigenvalues is a smooth manifold. To this end, charts are constructed explicitly,
and the dimension of~$\F^{p,q}$ is computed.
In Section~\ref{secriemann} it is shown that the Hilbert-Schmidt scalar product gives rise
to a Riemannian metric on~$\F^{p,q}$.
Section~\ref{secgauss} is devoted to the construction of corresponding Gaussian charts.
In Section~\ref{secgaugefix} our distinguished gauges are constructed, and
it is shown that the corresponding symmetric wave charts are indeed Gaussian charts.
Finally, in Section~\ref{secex} our constructions and results are illustrated for causal
fermion systems constructed from systems of Dirac wave functions in Minkowski space.

\section{Preliminaries on Causal Fermion Systems} \label{secprelim}
\subsection{From Quantum Mechanics to Causal Fermion Systems} \label{secquantumcfs}
This section is intended for readers who are not familiar with causal fermion systems.
Our presentation has similarities to other introductions (for example in~\cite[Section~2]{dice2014}, \cite[Section~1]{nrstg} or~\cite[Section~1.2]{cfs}, \cite[Section~4]{review}),
but it is streamlined towards clarifying the connection to local gauge symmetries in Minkowski space.

We begin in the setting of relativistic quantum mechanics in the presence of an external
classical electromagnetic field.
Let~$\scrM$ be Minkowski space and $\mu$ the natural volume measure thereon, i.e.~$d\mu = d^4x$ if $x=(x^0,x^1,x^2,x^3)$ is an inertial frame.
We consider Dirac wave functions in the presence of an external electromagnetic potential~$A$,
which satisfy the Dirac equation
\begin{align}
\label{DiracEq}
\big( i \gamma ^j \partial_j + \gamma^j A_j - m \big)\, \psi = 0 \:,
\end{align}
where~$m$ is the rest mass and~$\gamma^j$ are Dirac matrices in the Dirac representation.
On the Dirac solutions, we consider the usual scalar product
\begin{align}
\label{ScalProd}
( \psi | \phi )_t := 2 \pi \int_{t=\textrm{const}} (\overline \psi \gamma^0  \phi) (t,\vec x) \:d^3 x
\end{align}
(here~$\overline{\psi} = \psi^\dagger \gamma^0$ is the adjoint spinor, where the dagger denotes complex conjugation and transposition). If one evaluates~\eqref{ScalProd} for~$\phi=\psi$,
the integrand can be written as~$(\overline{\psi}\gamma^0\psi)(t,\vec{x}) = (\psi^\dagger \psi)(t,\vec{x})$,
having the interpretation as the probability density of the Dirac particle corresponding to~$\psi$
to be at time~$t$ at the position~$\vec{x}$. Due to current conservation, the integral in~\eqref{ScalProd} is time independent.

As already mentioned in the introduction, the above system is {\em{invariant under
local gauge transformations}}. More precisely, this means that
combining the transformation of the electromagnetic potential~\eqref{gauge1}
with the local phase transformation~\eqref{gauge2} of the wave functions,
the Dirac equation~\eqref{DiracEq} is preserved. Moreover, the scalar product~\eqref{ScalProd}
remains unchanged. Also, all observables (like local densities, momenta, etc.) are preserved.
In simple terms, one can say that the combined transformation~\eqref{gauge1} and~\eqref{gauge2}
does not change the physical content of the system.

Next, we choose an ensemble of Dirac solutions~$\psi_1, \ldots, \psi_f$.
For simplicity in presentation, we restrict attention to the case~$f<\infty$ of a finite number of
Dirac wave functions, which we assume to be continuous.
It is a central idea behind causal fermion systems to describe the physical system and
to formulate its dynamical equations purely in terms of the ensemble of wave
functions~$\psi_1, \ldots, \psi_f$.
Another idea is that the causal fermion system should encode the form of the wave functions
in a gauge-invariant way. To this end, we denote the complex vector space spanned by the
wave functions~$\psi_1, \ldots, \psi_f$ by~$\H$. On~$\H$ we consider the restriction of the
scalar product~\eqref{ScalProd}, i.e.\ $\la .|. \ra_\H := ( .|. )_t|_{\H \times \H}$.
Thus~$(\H, \la .|. \ra_\H)$ is an $f$-dimensional complex vector space.
Its vectors are represented by wave functions, which are defined only up to local phases
as described by the gauge transformation~\eqref{gauge2}.
For any spacetime point~$x \in \scrM$, we now introduce the sesquilinear form
\beq \label{bxdef}
b_x :  \H \times \H \rightarrow \C \:,\qquad b_x(\psi, \phi) = -(\overline \psi \phi) (x) \:,
\eeq
which maps two solutions of the Dirac equation to their inner product at $x$.
The sesquilinear form $b_x$ can be represented by a self-adjoint operator $F(x)$ on $\H$,
which is uniquely defined by the relations
\[ \la \psi | F(x) \phi \ra_\H =b_x(\psi,\phi) \qquad \text{for all~$\psi, \phi \in \H$}\:. \]
More concretely, in the basis~$(\psi_k)_{k = 1, \ldots,f}$ of~$\H$, the last relation can be written as
\begin{align} \label{Fdef}
\la \psi_i | F(x) \psi_j \ra_\H = - \big(\overline{\psi_i} \psi_j \big)(x) \:.
\end{align}
If the basis is orthonormal, the calculation
\[ F(x) \,\psi_j = \sum_{i=1}^f \la \psi_i | F(x) \psi_j \ra_\H\; \psi_i
= - \sum_{i=1}^f \big(\overline{\psi_i} \psi_j \big)(x)\; \psi_i \]
(where we used the completeness relation~$\phi = \sum_i \la \psi_i | \phi \ra\, \psi_i$),
shows that the operator~$F(x)$ has the matrix representation
\[ \big(F(x) \big)^i_j = - \big(\overline{\psi_i} \psi_j \big)(x) \:. \]
In physical terms, the matrix element~$-(\overline{\psi_i} \psi_j)(x)$ gives information on the correlation of the
wave functions~$\psi_i$ and~$\psi_j$ at the spacetime point~$x$.
Therefore, we refer to~$F(x)$ as the {\em{local correlation operator}} at~$x$.

Let us analyze the properties of $F(x)$. First of all, the calculation
\[ \la F(x) \,\psi \,|\, \phi \ra_\H = \overline{ \la \phi \,|\, F(x) \,\psi \,\ra_\H}
= -\overline{(\overline \phi \psi) (x)} = -(\overline \psi \phi) (x) = \la \psi \,|\, F(x) \,\phi \ra_\H \]
shows that the operator~$F(x)$ is self-adjoint
(where we denoted complex conjugation by a bar).
Furthermore, since the pointwise inner product $(\overline \psi \phi)(x)$ has signature $(2,2)$,
we know that~$b_x$ has signature $(p,q)$ with $p,q \leq 2$.
As a consequence, counting multiplicities, the operator $F(x)$ has at most two positive and at most two negative eigenvalues. It is useful to denote the set of all symmetric linear operators on~$\H$ which have rank at most
four and (counting multiplicities) have at most two positive and at most two negative eigenvalues by~$\F
\subset \Lin(\H)$. Then the local correlation operator~$F(x)$ is an element of~$\F$.

Constructing the operator $F(x) \in \F$ for every spacetime point $x \in \scrM$, we
obtain the {\em{local correlation map}}
\[ F : \scrM \rightarrow \F \:,\qquad x \mapsto F(x) \:. \]
This allows us to introduce a measure $\rho$ on $\F$ as follows. For any~$\Omega \subset \F$,
one takes the pre-image $F^{-1}(\Omega) \subset \scrM$ and computes its spacetime volume,
\[ \rho(\Omega) := \mu \big( F^{-1}(\Omega) \big) \:. \]
This gives rise to the so-called {\em{push-forward measure}} which in mathematics
is denoted by~$\rho = F_\ast \mu$. The $\rho$-measurable sets are defined as
the $\sigma$-algebra of all subsets of~$\F$ whose pre-image~$F^{-1}(\Omega)$
is $\mu$-measurable.

The resulting triple~$(\H, \F, \rho)$ is a causal fermion system
(for the abstract definition see Definition~\ref{defparticle} below).
Before going on, we make a few remarks on the above construction.
We first point out that the sesquilinear form~$b_x$ in~\eqref{bxdef}
and consequently also the operators~$F(x)$ are invariant under local
phase transformations~\eqref{gauge2}. Thus the causal fermion system is
defined in a manifestly gauge-invariant manner.
At first sight, this might seem to entail that local gauge freedom plays no role
in this approach. However, this view is too simple for the following reason:
Starting from a general causal fermion system~$(\H, \F, \rho)$, the vectors of the
Hilbert space are merely abstract vectors in the sense that, a-priori, they are not
represented by wave functions in spacetime. But one can construct
corresponding wave functions, the so-called {\em{physical wave functions}}.
This representation is canonical, but it is unique only up to local gauge transformations.
In this way, local gauge freedom again comes into play.

In order to make these concepts and their connections clearer, we
proceed by first giving the general definition of causal fermion systems
(Section~\ref{seccfs}). Then we explain how spacetime and the physical
wave functions arise (Section~\ref{secinherent}).
After restricting attention to the so-called regular setting (Section~\ref{secregular}),
we finally explain the resulting gauge freedom in more detail (Section~\ref{secgaugecfs}).

\subsection{Basic Definition} \label{seccfs}
We now give the abstract definition (for more details see for example~\cite[Section~1.1]{cfs}).
\begin{Def} \label{defparticle} (causal fermion system) {\em{ 
Given a separable complex Hilbert space~$\H$ with scalar product~$\la .|. \ra_\H$
and a parameter~$n \in \N$ (the {\em{``spin dimension''}}), we let~$\F \subset \Lin(\H)$ be the set of all
self-adjoint operators on~$\H$ of finite rank, which (counting multiplicities) have
at most~$n$ positive and at most~$n$ negative eigenvalues. On~$\F$ we are given
a positive measure~$\rho$ (defined on a $\sigma$-algebra of subsets of~$\F$), the so-called
{\em{universal measure}}. We refer to~$(\H, \F, \rho)$ as a {\em{causal fermion system}}.
}}
\end{Def}
The physical equations are formulated via a variational principle, the {\em{causal action principle}},
which we now introduce.

\subsection{A Few Inherent Structures} \label{secinherent}
{\em{Spacetime}}~$M$ is defined as the support of the universal measure,
\[ M := \text{supp}\, \rho \subset \F \]
(where the support is defined as the complement of the largest open set of measure zero).
For every~$x \in \F$ we define the
\beq \label{Sdef}
\text{\em{spin space}} \qquad S_x:= x(\H) \:;
\eeq
it is a subspace of~$\H$ of dimension at most~$2n$. 
On~$S_x$ we choose the inner product
\beq \label{ssp}
\Sl .|. \Sr_x \::\: S_x \times S_x \rightarrow \C \:, \qquad 
\Sl u | v \Sr_x = -\la u | x v \ra_\H \:,
\eeq
referred to as the {\em{spin inner product}}. It is an indefinite inner product
of signature~$(\p, \q)$ with~$\p, \q \leq n$.

A {\em{wave function}}~$\psi$ is defined as a function
which to every spacetime point~$x \in M$ associates a vector of the corresponding spin space,
\[ 
\psi \::\: M \rightarrow \H \qquad \text{with} \qquad \psi(x) \in S_x \quad \text{for all~$x \in M$}\:. \]
Every vector~$u \in \H$ gives rise to a corresponding wave function, referred to as the
{\em{physical wave function}}~$\psi^u$. It is defined by
\[ \psi^u(x) = \pi_x u \in S_x \:, \]
where~$\pi_x : \H \rightarrow S_x$ is the orthogonal projection in~$\H$
to the subspace~$S_x \subset \H$.
Finally, it is convenient to combine all the physical wave functions to an operator,
the so-called {\em{wave evaluation operator}}~$\Psi$ defined for any~$x \in \F$ by
\beq \label{Psixdef}
\Psi(x) \::\: \H \rightarrow S_x \:,\qquad u \mapsto \pi_x u \:.
\eeq
Then clearly, for every spacetime point~$x \in M$ and every~$u \in \H$,
\[ \Psi(x)\, u = \psi^u(x) \:. \]

In what follows, we shall often take adjoints of the above operators.
When doing so, one must be careful to work with the correct corresponding inner products.
In order to avoid confusion, we now explain in detail how this works.
The adjoint of~$\Psi(x)$ is defined formally by
\[ \Psi(x)^* \::\: S_x \rightarrow \H \:, \qquad
\Sl \phi \,|\, \Psi(x) \,u\Sr_x = \la \Psi(x)^* \,\phi \:|\: u \ra_\H \quad
\text{for all~$u \in \H$ and~$\phi \in S_x$}\:. \]
Note that on the left side of this equation the spin inner product appears.
Using its definition~\eqref{ssp}, we obtain the relation
\beq \label{cond2}
-\la \phi \,|\, X\, \Psi(x) \,u\ra_\H = \la \Psi(x)^* \,\phi \:|\: u \ra_\H \:,
\eeq
where we introduced the short notation
\beq \label{Xdef}
X := x|_{S_x} \::\: S_x \rightarrow S_x \:.
\eeq
Now we can take adjoints purely with respect to the Hilbert space scalar product.
Denoting those adjoints for clarity by a dagger, we can rewrite~\eqref{cond2} as
\beq \label{dagger}
-\la \Psi(x)^\dagger\,X \phi \,|\, u\ra_\H = \la \Psi(x)^* \,\phi \:|\: u \ra_\H \:,
\eeq
implying that
\beq \label{Psidagger}
\Psi(x)^* = -\Psi(x)^\dagger\,X \:.
\eeq
Adjoints of other operators can be computed similarly.
For a linear operator~$A \in \Lin(S_x)$, for example, the adjoint is defined by
\[ \Sl  \phi | A \tilde{\phi} \Sr_x = \Sl A^* \phi | \tilde{\phi} \Sr_x \qquad
\text{for all~$\phi, \tilde{\phi} \in S_x$}\:. \]
Using again the definition of the spin inner product~\eqref{ssp}, we can rewrite this equation as
\[ -\la \phi \,|\, X\,A \tilde{\phi} \ra_\H = -\la A^* \phi \,|\,X \tilde{\phi} \ra_\H \:, \]
and taking adjoints in the Hilbert space~$\H$ gives
\[ -\la X^{-1}\, A^\dagger\,X \phi \,|\,X \tilde{\phi} \ra_\H = -\la A^* \phi \,|\,X \tilde{\phi} \ra_\H \]
(note that the operator~$X$ is invertible because~$S_x$ is
by definition its image~\eqref{Sdef}). We thus obtain
\beq \label{Astar}
A^* = X^{-1}\, A^\dagger\,X \:.
\eeq

We now derive an identity which will be important later on (for an alternative derivation see~\cite[Lemma~1.1.3]{cfs}).
\begin{Lemma} For all~$x \in \F$,
\begin{align}
x &= -\Psi(x)^* \,\Psi(x) \label{xrep} \\
&= \Psi(x)^\dagger\,X \,\Psi(x)\:. \label{xrepdagger}
\end{align}
\end{Lemma}
\Proof Combining~\eqref{Psidagger} and~\eqref{Psixdef}, we obtain
\[ \Psi(x)^* \,\Psi(x) = -\Psi(x)^\dagger\,X\, \Psi(x) = -\pi_x^\dagger\,X\, \pi_x
= -\pi_x\,X\, \pi_x\:, \]
where in the last step we used that orthogonal projections are symmetric operators on~$\H$.
Using~\eqref{Xdef} gives~\eqref{xrep}. Rewriting this relation with the help of~\eqref{Psidagger}
gives~\eqref{xrepdagger}.
\QED

\subsection{Restriction to Regular Causal Fermion Systems} \label{secregular}
In the definition of causal fermion systems, the number of positive or negative eigenvalues
of the operators in~$\F$ can be strictly smaller than~$n$.
This is important because it makes~$\F$ a closed subspace of~$\Lin(\H)$ (with respect to the
norm topology), which in turn is crucial for the general existence results for minimizers
of the causal action principle (see~\cite{continuum} or~\cite{intro}).
However, in all physical examples in Minkowski space or in a Lorentzian spacetime,
all the operators in~$M$ do have exactly~$n$ positive and exactly~$n$ negative eigenvalues.
This motivates the following definition (see also~\cite[Definition~1.1.5]{cfs}).
\begin{Def} \label{defregular} {\em{
An operator~$x \in \F$ is said to be {\em{regular}} if it has the maximal possible rank,
i.e.~$\dim x(\H) = 2n$. Otherwise, the operator is called {\em{singular}}.
A causal fermion system is {\em{regular}} if all its spacetime points are regular.}}
\end{Def} \noindent
In what follows, we restrict attention to regular causal fermion systems. Moreover, it is
convenient to also restrict attention to all those operators in~$\F$ which are regular,
\beq \label{Freg}
\F^\text{reg} := \big\{ x \in \F \:|\: \text{$x$ is regular} \big\} \:.
\eeq
$\F^\text{reg}$ is a dense open subset of~$\F$ (again with respect to the
norm topology on~$\Lin(\H)$).
For notational convenience, in omit the superscript ``reg'' from now on. Thus, in what follows,
\beq \label{allreg}
\text{by} \quad \F \quad \text{we always mean} \quad \F^\text{reg}\:.
\eeq

\subsection{Local Gauge Invariance and Gauge Transformations} \label{secgaugecfs}
The setting of causal fermion systems is {\em{gauge invariant}} in the following sense
(see also~\cite[Section~1.3]{cfs}):
In order to represent the wave functions in components, one must work with
basis representations of the spin spaces. To this end, we choose
a pseudo-orthonormal basis~$(\mathfrak{e}_\alpha(x))_{\alpha=1,\ldots, 2n}$
of every spin space~$(S_x, \Sl .|. \Sr_x)$, i.e.\
\[ \Sl \mathfrak{e}_\alpha(x) | \mathfrak{e}_\beta(x) \Sr_x = s_\alpha\: \delta^\alpha_\beta \]
with~$s_1=\ldots=s_n=1$ and~$s_{n+1}=\ldots=s_{2n}=-1$.
Then a wave function~$\psi$ can be represented as
\beq \label{component}
\psi(x) = \sum_{\alpha=1}^{2n} \psi^\alpha(x)\: \mathfrak{e}_\alpha(x)
\eeq
with component functions~$\psi^1, \ldots, \psi^{2n}$.
The freedom in choosing the basis~$(\mathfrak{e}_\alpha)$ is described by the
group~$\U(n,n)$ of unitary transformations with respect to an inner product of signature~$(n,n)$.
This gives rise to the transformations
\beq \label{locgauge}
\mathfrak{e}_\alpha(x) \rightarrow \sum_{\beta=1}^{2n} U^{-1}(x)^\beta_\alpha\;
\mathfrak{e}_\beta(x) \qquad \text{and} \qquad
\psi^\alpha(x) \rightarrow  \sum_{\beta=1}^{2n} U(x)^\alpha_\beta\: \psi^\beta(x)
\eeq
with $U \in \U(n,n)$.
As the basis~$(\mathfrak{e}_\alpha)$ can be chosen independently at each spacetime point,
one obtains {\em{local gauge transformations}} of the wave functions,
where the gauge group is determined to be the isometry group of the spin inner product.
The causal action is
{\em{gauge invariant}} in the sense that it does not depend on the choice of spinor bases.

We finally explain how this notion of gauge invariance is related to the
gauge freedom in general relativity and in the standard model.
It is important to observe that in our approach, the gauge group is determined
by the spin dimension: it is the group~$\U(n,n)$ of all unitary transformations of the spin space.
This group contains the group $\U(1)$ of electrodynamics, and the corresponding gauge
transformations~\eqref{locgauge} give the local phase transformations~\eqref{gauge2}.
In the case of spin dimension two, the group~$\U(2,2)$ also contains a covering of the
Lorentz group, making it possible to describe general relativity as a gauge theory~\cite{u22}.
If the spin dimension is larger, there are mechanisms which give rise to constraints,
leading to smaller effective gauge groups (for details see~\cite[Chapters~3-5]{cfs}).
In order to understand how massive gauge fields (like the $W$- or $Z$-bosons in the standard model)
come up, one must keep into account that left-handed and axial gauge potentials
do not correspond to gauge transformations of the form~\eqref{locgauge} because the
resulting local gauge transformations are not unitary with respect to the spin inner product.
This gives rise to a mass term without contradicting local gauge invariance.
This point and the connection to spontaneous symmetry breaking is explained in detail
in~\cite[\S3.6.2 and \S3.8.5]{cfs}. For the purpose of this paper, these effects are
not relevant because we mainly restrict our attention to the $\U(1)$ gauge transformations of
electromagnetism.

\section{A Smooth Manifold Structure of~$\F$} \label{secmanifold}
We now assume for technical simplicity that the Hilbert space~$\H$ is finite dimensional,
\[ \dim \H =: f < \infty \:. \]
For the sake of larger generality, instead of~$\F$ we consider operators
with different numbers of positive and negative eigenvalues. These operators are of importance
in view of topological and Riemannian fermion systems as introduced and analyzed in~\cite{topology}.
\begin{Def} We let~$\F^{p,q}$ be the set of all symmetric linear operators on~$\H$ of rank~$p+q$,
which (counting multiplicities) have~$p$ positive and~$q$ negative eigenvalues.
\end{Def} \noindent
Clearly, setting~$p=q=n$, we obtain the set~$\F$ (or, more precisely, the set~$\F^\text{reg}$;
see~\eqref{Freg} and~\eqref{allreg}).

\begin{Thm} \label{thmpq}
The set~$\F^{p,q}$ is a smooth manifold of dimension
\[ \dim \F^{p,q} = 2 \,(p+q)\,f - (p+q)^2\:. \]
\end{Thm}
\Proof Let~$x \in \F^{p,q}$. We denote its image by~$I \subset \H$ and set~$J=I^\perp$
(where the orthogonal complement is taken with respect to the scalar product on~$\H$).
Using a block matrix representation in~$\H = I \oplus J$, the operator~$x$ has the representation
\[ 
x = \begin{pmatrix} X & 0 \\ 0 & 0 \end{pmatrix} \:. \]
We now let~$A$ be symmetric linear operator on~$I$. By choosing its norm sufficiently small, we
can arrange that the operator~$X+A$ has again~$p$ positive and~$q$ negative eigenvalues.
In particular, this operator is invertible.
Next, we choose a linear operator~$B : J \rightarrow I$. We form the operator
\begin{align}
M &:= \begin{pmatrix} \1 & 0 \\ B^\dagger (X+A)^{-1} & \1 \end{pmatrix}
\begin{pmatrix} X+A & 0 \\ 0 & 0 \end{pmatrix}
\begin{pmatrix} \1 & (X+A)^{-1} B \\ 0 & \1 \end{pmatrix} \label{Mform0} \\
&\:= \begin{pmatrix} X+A & B \\ B^\dagger & B^\dagger (X+A)^{-1} B \end{pmatrix} \label{Mform1}
\end{align}
(where for clarity the dagger again denotes the adjoint with respect to the scalar product induced
from~$\la .|. \ra_\H$; see~\eqref{dagger}).
This operator is symmetric and has again~$p$ positive and~$q$ negative eigenvalues.
Thus for sufficiently small~$\varepsilon$ we obtain the mapping
\[ \Lambda \::\: \big(\Symm(I) \oplus \Lin(I,J) \big) \cap B_\varepsilon(0) \rightarrow \F^{p,q}\:,\qquad
(A,B) \mapsto M \]
(where~$\Symm(I)$ denotes the linear operators on~$I$ which are symmetric with
respect to the induced scalar product~$\la .|. \ra_\H|_{I \times I}$). Let us verify that (again for
sufficiently small~$\varepsilon$) this mapping is a homeomorphism to an open neighborhood
of~$x \in \F^{p,q}$. It is obvious from~\eqref{Mform1} that~$\Lambda$ is injective. In order to
verify that it maps to an open neighborhood of~$x$, we let~$y \in F^{p,q}$ with~$\|x-y\| < \delta$
(with~$\delta>0$ to be specified below). Diagonalizing~$y$ with a unitary operator~$U$, we obtain
the block matrix representation
\[ y = \begin{pmatrix} U_{11} & U_{12} \\ U_{21} & U_{22} \end{pmatrix}
\begin{pmatrix} X+C & 0 \\ 0 & 0 \end{pmatrix}
\begin{pmatrix} U_{11}^\dagger & U_{21}^\dagger \\ U_{12}^\dagger & U_{22}^\dagger \end{pmatrix} \:, \]
where~$C$ is a symmetric linear operator on~$I$.
In the limit~$y \rightarrow x$, the image of~$y$ converges to the image of~$x$,
implying that the operator~$U_{11}$ becomes unitary. Therefore, for sufficiently small~$\delta>0$,
the operator~$U_{11}$ is invertible, giving rise to the representation
\[ y = \begin{pmatrix} \1 & 0 \\ U_{21} \,U_{11}^{-1} & \1 \end{pmatrix}
\begin{pmatrix} U_{11}\, (X+C)\, U_{11}^\dagger & 0 \\ 0 & 0 \end{pmatrix}
\begin{pmatrix} \1 & (U_{11}^\dagger)^{-1}\, U_{21}^\dagger \\ 0 & \1 \end{pmatrix} \:. \]
This is indeed of the form~\eqref{Mform0}, and one can even read off~$A$ and~$B$,
\begin{align*}
A &= U_{11}\, (X+C)\, U_{11}^\dagger - X \\
B &= \big( U_{11}\, (X+C)\, U_{11}^\dagger \big)\, \big( U_{11}^\dagger)^{-1}\, U_{21}^\dagger \big)\:.
\end{align*}
We conclude that~$\Lambda$ is a bijection to an open neighborhood of~$x \in \F^{p,q}$.
The continuity of~$\Lambda$ and of its inverse are obvious. We have thus constructed a chart
on~$\F^{p,q}$ around~$x$.

Performing the above construction around every point of~$\F^{p,q}$ gives an atlas. By direct computation
one verifies that the transition maps are smooth. We conclude that, with this atlas,
$\F^{p,q}$ is indeed a smooth manifold.

We finally determine the dimension of~$\F^{p,q}$.
The linear operator~$B$ is represented by a~$(p+q) \times (f-p-q)$-matrix,
giving rise to~$2 (p+q)(f-p-q)$ real degrees of freedom. The symmetric linear operator~$A$, on the other
hand, is represented by a Hermitian $(p+q) \times (p+q)$-matrix, described by~$(p+q)^2$ real parameters.
Adding these dimensions concludes the proof.
\QED

From now on, we always restrict attention to the case~$p=q=n$ of causal fermion systems.

\section{A Riemannian Metric on~$\F$} \label{secriemann}
We finally introduce another inherent structure which has not been used so far
and which seems useful in the context of gauge fixing: a Riemannian metric on~$\F$.
As in the previous section, we assume for technical simplicity that~$\H$ is finite-dimensional.
Then on~$\F$ the Hilbert-Schmidt norm gives rise to a distance function
\beq \label{ddef}
d \::\: \F \times \F \rightarrow \R^+_0\:,\qquad
d(x,y) = \|x-y\|_{\text{\tiny{HS}}} := \sqrt{\tr \big((x-y)^2 \big)}
\eeq
(note that the existence of the trace would not be an issue even in the infinite-dimensional setting
because all operators in~$\F$ have finite rank).
The square of this distance function is smooth. Moreover, its first derivative vanishes
on the diagonal, i.e.~$D(d(x,.)^2)|_x=0$.
Therefore, taking its quadratic Taylor expansion
about a point~$x \in M$ gives a scalar product on~$T_x\F$, i.e.
\beq \label{hxdef}
h_x \::\: T_x\F \times T_x\F \rightarrow \R \:,\qquad
h_x(u,v) = \tr(u v) \:.
\eeq
Clearly, this mapping depends smoothly on~$x$ and thus defines a 
{\em{Riemannian metric}} on~$\F$.

\section{Gaussian Charts} \label{secgauss}
Specializing to the case~$p=q=n$, in the proof of Theorem~\ref{thmpq} we constructed a
local parametrization of~$\F$ given by
\beq \label{Lambda}
\begin{split}
\Lambda \::\: \big(\Symm(I) &\oplus \Lin(I,J) \big) \cap B_\varepsilon(0) \rightarrow \F\:,\\
(A,B) &\mapsto M = \begin{pmatrix} X+A & B \\ B^\dagger & B^\dagger (X+A)^{-1} B \end{pmatrix}\:.
\end{split}
\eeq
The image of this mapping is an open neighborhood
of~$x \in \F$ which we denote by~$U$. Then the inverse of~$\Lambda$ defines a chart
\beq \label{Lamchart}
\phi_x := \Lambda^{-1} \::\: U \subset \F \rightarrow \Symm(I) \oplus \Lin(I,J) \:.
\eeq

\begin{Thm} \label{thmgauss}
The chart~$(\phi_x, U)$ in~\eqref{Lamchart} is a Gaussian coordinate system
about the point~$x \in U$ with respect to the Riemannian metric~$h$ on~$\F$ (see~\eqref{hxdef}).
\end{Thm}
\Proof In the chart~$\phi_x$, we describe points of~$\F$ by pairs
\[ (A,B) \in \big(\Symm(I) \oplus \Lin(I,J) \big) \cap B_\varepsilon(0) \:. \]
Expanding the mapping~$\Lambda$ in a Taylor series about the origin,
there is a nonlinearity only in the lower right block matrix entry,
\[ \Lambda(tA, tB) = \begin{pmatrix} X & 0 \\ 0 & 0 \end{pmatrix}
+ t \: \begin{pmatrix} A & B \\ B^\dagger & 0 \end{pmatrix}
+ \begin{pmatrix} 0 & 0 \\ 0 & \O\big( t^2 \big) \end{pmatrix} \:. \]
Hence the distance function~\eqref{ddef} has the expansion
\begin{align}
d\big( &(tA, tB),\, (t \tilde{A}, t \tilde{B}) \big)^2
= \tr \Big\{ \big[ \Lambda(tA, tB) -  \Lambda(t \tilde{A}, t \tilde{B}) \big]^2 \Big\} \notag \\
&= \tr \bigg\{ \bigg[ t \: \begin{pmatrix} A-\tilde{A} & B-\tilde{B} \\ B^\dagger-\tilde{B}^\dagger & 0 \end{pmatrix}
+ \begin{pmatrix} 0 & 0 \\ 0 & \O\big( t^2 \big) \end{pmatrix} \bigg]^2 \bigg\} \notag \\
&= t^2 \: \tr \bigg\{ \begin{pmatrix} A-\tilde{A} & B-\tilde{B} \\ B^\dagger-\tilde{B}^\dagger & 0 \end{pmatrix}^2 \bigg\}
+ \O\big( t^4 \big) \:, \label{gauss}
\end{align}
where in the last step we made use of the cubic term in~$t$ is trace-free, because
\[ \tr \bigg\{ \begin{pmatrix} A-\tilde{A} & B-\tilde{B} \\ B^\dagger-\tilde{B}^\dagger & 0 \end{pmatrix}
\begin{pmatrix} 0 & 0 \\ 0 & (*) \end{pmatrix} \bigg\} 
= \tr \begin{pmatrix} 0 & (B-\tilde{B})\,(*) \\ 0 & 0 \end{pmatrix} = 0 \]
(where the star stands for an arbitrary block matrix entry).

The formula~\eqref{gauss} shows that, in our coordinates, the Riemannian metric is constant
up to contributions of order~$\O(t^2)$. Therefore, the coordinates are indeed Gaussian.
\QED

\section{Gauges and Gauge Fixing} \label{secgaugefix}
We saw in Section~\ref{secgaugecfs} that the vectors in~$\H$ can be represented by $(2n)$-component
wave functions in spacetime~\eqref{component}, unique up to local gauge transformations~\eqref{locgauge}.
In order to clarify the mathematical structures, it is useful to choose an inner product space~$(V, \la .|. \ra)$
of signature~$(n,n)$ with pseudo-orthonormal basis~$(\mathfrak{f}_1, \ldots, \mathfrak{f}_{2n})$.
Then one can regard the~$\psi^\alpha(x)$ in~\eqref{component} as component
functions of vectors in~$V$,
\[ \sum_{\alpha=1}^{2n} \psi^\alpha(x)\: \mathfrak{f}_\alpha \;\in\; V \:, \]
We thus obtain a representation of~$\H$ as $V$-valued functions in spacetime.
The only condition to fulfill is that at each spacetime point~$x$, the resulting local correlation operator
must coincide with the operator~$x \in \F$. This leads us to the following notion:
\begin{Def} \label{defgauge} Let~$(V, \Sl .|. \Sr)$ be an indefinite inner product space of signature~$(n,n)$.
Moreover, let~$\Omega \subset \F$ be an open spacetime region. A mapping
\[ \Psi^\Omega_V \::\: \Omega \rightarrow \Lin(\H,V) \]
is called a {\bf{gauge}} in~$\Omega$ if
\[ x = - \big(\Psi^\Omega_V(x) \big)^* \big(\Psi^\Omega_V(x) \big) \qquad \text{for all~$x \in \Omega$}\:. \]
\end{Def} \noindent
Here the adjoint is to be taken with respect to the corresponding inner products, i.e.\
\[ \Sl \phi \,|\, \Psi^\Omega_V(x) \,u \Sr = \la \big(\Psi^\Omega_V(x) \big)^* \phi \,|\, u \ra_\H 
\qquad \text{for all~$\phi \in V$ and~$u \in \H$}\:. \]
We remark that the concept of defining a gauge as a representation of Hilbert space vectors
as wave functions goes back to~\cite[Definition~2.1]{gauge}.

In order to see that gauges exist, one can proceed as follows.
Given~$\Omega$, for every~$y \in \Omega$ one chooses a  unitary mapping
\beq \label{SViso}
U_y \::\: S_y \rightarrow V
\eeq
(such a unitary mapping exists because~$V$ and~$S_y$ have the same signature).
Then the mapping
\beq \label{Psigauge}
\Psi^\Omega_V(y) := U_y \,\Psi(y)
\eeq
is indeed a gauge because
\[ -\big(\Psi^\Omega_V(y) \big)^* \big(\Psi^\Omega_V(y) \big)
= -\big(\Psi(y) \big)^* \big(\Psi(y) \big) = y \]
(where in the first step we used that~$U_y$ is unitary, and in the second step we applied~\eqref{xrep}).
In this construction, the local gauge freedom corresponds to the freedom in choosing the
isomorphisms~\eqref{SViso} between the spin spaces and~$V$.

{\em{Fixing the gauge}} amounts to constructing distinguished gauges.
Our general procedure is outlined as follows. Given~$x \in \F$, we want to construct
a distinguished gauge in an open neighborhood~$\Omega \subset \F$ of~$x$.
To this end, we want to construct a distinguished mapping
\beq \label{phidef}
\phi \::\: \Omega \rightarrow \Lin(\H, S_x)
\eeq
with the property that the local correlation operator corresponding to~$\phi(y)$
agrees with~$y$, i.e.
\[ y = - \phi(y)^* \phi(y) \qquad \text{for all~$y \in \Omega$}\:. \]
Next, we choose a unitary operator~$U_x$ from~$S_x$ to~$V$. Then the mapping
\beq \label{Psifix}
\Psi^\Omega_V \::\: \Omega \rightarrow \Lin(\H,V) \:,\qquad \Psi^\Omega_V(y) := U_x \, \phi(y)
\eeq
is a gauge. This construction is illustrated in the following diagram:
\begin{center}
\begin{tikzpicture}
\matrix (m) [matrix of math nodes,row sep=1.5cm,column sep=2cm,minimum width=2cm]
{
y \in \Omega \subset \F &	\phi(y) \in \Lin(\H, S_x)  \\
	 & U_x \, \phi(y) \in \Lin(\H, V) \\};
\path[-stealth]
(m-1-1) edge node [above] {$\phi$} (m-1-2)
node [above] {} (m-2-2)
(m-1-2) edge node [right] {$U_x : S_x \rightarrow V$} (m-2-2)
(m-1-1) edge node [above] {$\;\;\;\Psi^\Omega_V$} (m-2-2)
;
\end{tikzpicture}
\end{center}
Note that, in contrast to the construction~\eqref{Psigauge}, which involves the freedom
in choosing a unitary operator~$U_y$ at every~$y \in \Omega$,
the gauge~\eqref{Psifix} involves only one unitary operator~$U_x$.
In this way, the local gauge freedom has been fixed up to {\em{global}} gauge transformations.

The mapping~$\phi$ in~\eqref{phidef} has a simple interpretation as
``using wave functions as coordinates.'' Indeed, given~$u \in \H$,
the vector~$\phi(y) u \in S_x$ can be regarded as the physical wave function at
the spacetime point~$y$, however in a gauge where all the spin spaces are identified with~$S_x$.
This idea will become clearer in the next sections, when we use the wave evaluation operator~$\Psi$
for the construction of~$\phi$.
Due to the local gauge freedom, the idea of ``using wave functions as coordinates''
can be realized only after invoking a gauge fixing procedure.
We first introduce this gauge fixing by hand (Section~\ref{seccwc})
and justify it afterward by analyzing the Gaussian charts of Section~\ref{secgauss}
(Section~\ref{secgwc}).

\subsection{Symmetric Wave Charts} \label{seccwc}
By varying the wave evaluation operator, we obtain
a mapping
\beq \label{Rdef}
R \::\: W \subset \Lin(\H, S_x) \rightarrow \F\:,\qquad \psi \mapsto -\psi^* \psi \:,
\eeq
where~$W$ is an open neighborhood of~$\Psi(x)$ which is chosen so small that
all the operators in the image of~$R$ have~$n$ positive and~$n$ negative eigenvalues
(a similar construction which in addition arranges a constant trace is considered in see~\cite[Section~6.2]{perturb}).
Since every operator in~$\F$ can be realized as its own local correlation operator~\eqref{xrepdagger}
and all the spin spaces are isomorphic, it is obvious that the image of~$R$ contains an open neighborhood
of~$x \in \F$. However, the operator~$R$ has a kernel. In order to describe this kernel systematically, it is
convenient to again decompose the Hilbert space into the direct sum
\beq \label{dirsum}
\H = I \oplus J \qquad \text{with} \qquad I:=S_x,\; J:= (S_x)^\perp \:,
\qquad W \ni \psi = \psi_I + \psi_J\:.
\eeq
For clarity, we point out that~$I$ always denotes the Hilbert space with the induced scalar
product~$\la .|. \ra_\H|_{I \times I}$. Thus~$I$ and~$S_x$ coincide as complex vector spaces.
However, the spin space~$S_x$ is not a Hilbert space but an indefinite inner product space,
endowed with the spin inner product~$\Sl .|. \Sr_x := \la .|x.\ra_\H$.

The direct sum decomposition~\eqref{dirsum}
 gives rise to a corresponding decomposition of the linear operators,
\[ \Lin(\H, S_x) = \Lin(I, S_x) \oplus \Lin(J, S_x)\:. \]
We again point out that the space~$\Lin(I, S_x)$ coincides as a vector space with~$\Lin(I)$,
and~$\Lin(J,S_x)$ coincides with~$\Lin(J, I)$. However, when taking adjoints,
one must be careful to take the correct inner products.
Possibly by choosing~$W$ smaller, we can arrange that the operators~$\psi_I \in \Lin(I,I)$ are
all invertible, so that~$R$ becomes a mapping
\[ 
R \::\: W \subset \text{GL}(S_x) \oplus \Lin(J, S_x) \rightarrow \F \]
(where we canonically identified~$\Lin(I, S_x)$ with~$\Lin(S_x)$).
The gauge freedom becomes manifest in the fact that the mapping~$R$ has a non-trivial kernel:
\begin{Lemma} $R$ is injective up to gauge transformations in~$\U(S_x)$, meaning that
\beq \label{injective}
R\big( \psi_I, \psi_J) = R\big( \tilde{\psi}_I, \tilde{\psi}_J) \quad \Longleftrightarrow \quad
\exists \;U \in \U(S_x) \text{ with }
\tilde{\psi}_I = U \psi_I \text{ and } \tilde{\psi}_J = U \psi_J
\eeq
(where~$\U(S_x)$ are the unitary operators with respect to the spin inner product).
\end{Lemma}
\Proof Unitarity with respect to the spin inner product is defined by
\[ \Sl U \phi | U \tilde{\phi} \Sr_x = \Sl \phi | \tilde{\phi} \Sr_x \qquad
\text{for all~$\phi, \tilde{\phi} \in S_x$}\:. \]
Using~\eqref{Astar}, unitarity can be written more explicitly as the conditions
\beq \label{unitcond}
X^{-1}\, U^\dagger\, X = U^{-1} \qquad \text{or} \qquad U^\dagger\, X = X\, U^{-1} \:.
\eeq

Using a block matrix notation in the direct sum decomposition~\eqref{dirsum}, we have
\[ R\big( \psi_I, \psi_J) = -\begin{pmatrix} \psi_I^\dagger \,X\, \psi_I & \psi_I^\dagger \,X\, \psi_J \\[0.3em]
\psi_J^\dagger \,X\, \psi_I & \psi_J^\dagger \,X\, \psi_J \end{pmatrix} \:. \]
Hence the condition~$R\big( \psi_I, \psi_J) = R\big( \tilde{\psi}_I, \tilde{\psi}_J)$ is equivalent to the
three equations
\begin{align}
\psi_I^\dagger \,X\, \psi_I &= \tilde{\psi}_I^\dagger \,X\, \tilde{\psi}_I \label{c1} \\
\psi_I^\dagger \,X\, \psi_J &= \tilde{\psi}_I^\dagger \,X\, \tilde{\psi}_J \label{c2} \\
\psi_J^\dagger \,X\, \psi_J &= \tilde{\psi}_J^\dagger \,X\, \tilde{\psi}_J \:. \label{c3}
\end{align}
Since~$\psi_I$ and~$\tilde{\psi}_I$ are invertible operators, we can write
\beq \label{psiU} \tilde{\psi}_I = U \psi_I \eeq
with an invertible operator~$U$ on~$I$. Multiplying~\eqref{c1} from the left by the inverse of~$\psi_I^\dagger$
and from the right by the inverse of~$\psi_I$, we obtain the condition~$X=U^\dagger X U$.
Comparing with~\eqref{unitcond}, we conclude that~$U \in \U(S_x)$ is unitary with respect to the spin inner product.
Substituting~\eqref{psiU} in~\eqref{c2} and using that~$U$ is unitary with respect to the spin inner product,
we obtain the equivalent condition
\[ \psi_I^\dagger \,X\, \psi_J = \psi_I^\dagger\,U^\dagger \,X\, \tilde{\psi}_J = \psi_I^\dagger \,X\, U^{-1} \,\tilde{\psi}_J \:, \]
and multiplying from the left by~$U\,X^{-1}\, (\psi_I^\dagger)^{-1}$ gives the last identity in~\eqref{injective}.
If the relations on the right side of~\eqref{injective} hold, then the condition~\eqref{c3} is also satisfied.
This concludes the proof.
\QED

After these preparations, we can explain our method for fixing the gauge.
The gauge freedom~\eqref{injective} means that both~$\psi_I$ and~$\psi_J$ can be
multiplied by an arbitrary unitary operator~$U \in \U(S_x)$.
Using this freedom, we can arrange that
\beq \label{symmcond}
\psi_I \in \Symm(S_x)
\eeq
becomes a {\em{symmetric}} operator. This method indeed fixes the local gauge freedom completely,
as we shall now work out.
Before beginning, we remark that, at present, our procedure is motivated only by the fact that it works and
is canonical. A deeper justification will be given in connection with the Gaussian charts in Section~\ref{secgwc} below.

We begin with a preparatory lemma.
\begin{Lemma} {\bf{(unique polar decomposition)}} \label{lemmapolar}
Let~$(V, \Sl.|.\Sr)$ be a (finite-dimen\-sio\-nal) indefinite inner product space.
Then there is an open neighborhood~$W$ of~$\1 \in \Lin(V)$ such that every operator~$A \in W$
has a unique polar decomposition
\beq \label{Wcomp}
A = U\, S \qquad \text{with} \qquad U \in \U(V) \text{ and } S \in \Symm(V) \cap W \:.
\eeq
\end{Lemma}
\Proof Writing~$A=\1+\Delta A$, it follows that
\[ B:= A^* A = (\1+\Delta A)^* (\1+\Delta A) = \1 + \Delta B \]
with~$\Delta B = (\Delta A)^* + (\Delta A) + (\Delta A)^* (\Delta A)$.
If the neighborhood of~$W$ is chosen sufficiently small, the spectral calculus for~$B$
is well-defined as a power expansion in~$\Delta B$. In particular, the series
\begin{align*}
B^\frac{1}{2} &:= 1+\frac{\Delta B}{2} -\sum_{n=2}^\infty \frac{1}{2^n}\: \frac{(2n-3)!}{n!\, (2n-4)!!} \;
\big( -\Delta B \big)^n \:, \\
B^{-\frac{1}{2}} &:= 1 + \sum_{n=1}^\infty \frac{1}{2^n}\: \frac{(2n-1)!}{n!\, (2n-2)!!} \; \big( -\Delta B \big)^n
\end{align*}
converge absolutely (since~$V$ is finite-dimensional, all norms on~$V$ are equivalent,
defining a unique topology on~$V$).
This makes it possible to form the polar decomposition~\eqref{Wcomp}
with the standard formulas
\[ U = A\, B^{-\frac{1}{2}} \qquad \text{and} \qquad S = B^\frac{1}{2}\:. \]
Here the operator~$U$ is indeed unitary because
\[ U U^* = \big( A\, B^{-\frac{1}{2}} \big) \big( B^{-\frac{1}{2}}\, A^* \big) = A B^{-1} A^*
= A\, \big( A^* A \big)^{-1}\, A^* = A\, A^{-1}\, \big(A^*\big)^{-1}\, A^* = \1 \:. \]

It remains to show uniqueness. To this end, we consider two polar decompositions,
\beq \label{AUS}
A = U\,S = \tilde{U}\, \tilde{S}\:.
\eeq
Using that~$U, \tilde{U}$ are unitary and~$S, \tilde{S}$ are symmetric, we obtain
\beq \label{S2}
A^* A = S^2 = \tilde{S}^2 \:.
\eeq
Choosing~$W$ sufficiently small, the square root is again well-defined as a power series, i.e.
\[ S = \sqrt{S^2} \qquad \text{and} \qquad \tilde{S} = \sqrt{\tilde{S}^2} \:. \]
Hence~\eqref{S2} implies that~$S=\tilde{S}$. As a consequence, it follows from~\eqref{AUS}
that~$U=\tilde{U}$, completing the proof.
\QED

We now apply the previous lemma to the mapping~$\psi_I$:
\begin{Lemma} There is an open neighborhood~$W$ of~$\Psi(x) \in \Lin(\H, S_x)$
such that for every~$\psi \in \Omega$, the operator~$\psi_I : S_x \rightarrow S_x$ has a unique polar
decomposition of the form
\beq \label{psiIsymm}
\psi_I = U\, S \qquad \text{with} \qquad U \in \U(S_x) \text{ and } S \in \Symm(S_x) \cap \pi_x \Omega|_{S_x} \:.
\eeq
\end{Lemma}
\Proof Since
\[ \Psi(x)_I = \1_I\:, \qquad \Psi(x)_J = 0 \:, \]
we can write~$\psi \in \Omega$ as
\[ \psi_I = \1_I + \Delta \psi_I \:\qquad \psi_J = \Delta \psi_J \:, \]
where~$\Delta \psi$ is sufficiently small.
Regarding the operator~$\psi_I$ as an endomorphism of~$S_x$, we can apply Lemma~\ref{lemmapolar}
with~$V=S_x$ to conclude that this operator has a unique polar decomposition of the form~\eqref{psiIsymm}.
\QED

We now use the unitary operator~$U$ in~\eqref{psiIsymm} to perform a gauge transformation~\eqref{Wcomp}.
Due to the uniqueness of the construction, we thus obtain a chart.
We have thus proved the following theorem:
\begin{Thm} \label{thmchart1}
For every~$x \in \F$ there is an open neighborhood~$W$ of
\[ (\1, 0) \in \Symm(S_x) \oplus \Lin(J, S_x) \]
such that
\[ R \::\: W \subset \Symm(S_x) \oplus \Lin(J, S_x) \rightarrow \F \:,\qquad \psi \mapsto -\psi^* \psi \]
is a local parametrization of~$\F$ around~$x$. Its inverse
\beq \label{phichart}
\phi := \big( R|_\Omega \big)^{-1} \::\: \Omega \subset \F \rightarrow \Symm(S_x) \oplus \Lin(J, S_x)
\eeq
with~$\Omega:= R(W) \subset \F$ is a local chart of~$\F$.
\end{Thm} \noindent
The chart~$(\phi, \Omega)$ is referred to as the {\em{symmetric wave chart}} about the point~$x \in M$.

We finally bring the symmetric wave chart into a more explicit form:
\begin{Prp} \label{prpphi1}
Choosing the open set~$\Omega$ sufficiently small, the symmetric wave chart~$\phi$ in~\eqref{phichart}
takes the form
\[ \phi(y) = \big( P(x,x)^{-1}\, A_{xy}\, P(x,x)^{-1} \big)^{-\frac{1}{2}}\, P(x,x)^{-1}\, P(x,y) \:
\Psi(y) \:, \]
where~$A_{xy} := P(x,y)\, P(y,x)$ is the closed chain.
\end{Prp}
\Proof The operator~$R$ introduced in~\eqref{Rdef} has the property that
\[ -\Psi(y)^* \Psi(y) = y \overset{!}{=} R(\psi) = -\psi^* \psi \:. \]
As a consequence, $\psi$ differs from~$\Psi(y)$ by a unitary mapping, i.e.
\[ \psi = U_{x,y}\, \Psi(y) \::\: \H \rightarrow S_x \qquad \text{with} \qquad U_{x,y} \in \U(S_y, S_x)\:. \]
We thus obtain the ansatz
\beq \label{phiansatz}
\phi(y) = U_{x,y}\, \Psi(y) \:,
\eeq
where we must choose~$U_{x,y}$ such that the restriction~$\phi(y)|_{S_x} : S_x \rightarrow S_x$ is
symmetric. Let us evaluate what this condition means: First, it is convenient to express the operators in
terms of the kernel of the fermionic projector,
\[ \Psi(y) |_{S_x} = \pi_y\, \pi_x |_{S_x}
= \pi_y\, x\, X^{-1} \,\pi_x \big|_{S_x} = P(y,x)\, P(x,x)^{-1} \big|_{S_x}  \]
(where we used again the notation~\eqref{Xdef}).
Next, we form a polar decomposition of the obtained operator
\[ B := P(y,x)\, P(x,x)^{-1} \big|_{S_x}\::\:  S_x \rightarrow S_y \:. \]
This gives
\[ \phi(y)|_{S_x} = U_{x,y}\, \Psi(y)|_{S_x}
= U_{x,y}\, B|_{S_x}
= U_{x,y}\, \big( B\, (B^* B)^{-\frac{1}{2}} \big)\,(B^* B)^{\frac{1}{2}} \big|_{S_x} \:. \]
Therefore, we can choose~$U_{xy}$ as
\begin{align}
U_{x,y} &= \big( B\, (B^* B)^{-\frac{1}{2}} \big)^{-1} = (B^* B)^{\frac{1}{2}}\, B^{-1}
= (B^* B)^{\frac{1}{2}}\, (B^*B)^{-1}\, B^* =  (B^* B)^{-\frac{1}{2}}\, B^* \notag \\
&= \big( P(x,x)^{-1}\, A_{xy}\, P(x,x)^{-1} \big)^{-\frac{1}{2}}\, P(x,x)^{-1}\, P(x,y) \:. \label{Uxyex}
\end{align}
This operator is indeed unitary because
\begin{align*}
U_{x,y} \, (U_{x,y}^*)&=
\big( P(x,x)^{-1}\, A_{xy}\, P(x,x)^{-1} \big)^{-\frac{1}{2}}\, P(x,x)^{-1}\, P(x,y)  \\
&\qquad \times P(y,x)\, P(x,x)^{-1} \big( P(x,x)^{-1}\, A_{xy}\, P(x,x)^{-1} \big)^{-\frac{1}{2}} \\
&= \big( P(x,x)^{-1}\, A_{xy}\, P(x,x)^{-1} \big)^{-\frac{1}{2}} \\
&\qquad \times \big( P(x,x)^{-1}\, A_{xy}\, P(x,x)^{-1} \big)
\big( P(x,x)^{-1}\, A_{xy}\, P(x,x)^{-1} \big)^{-\frac{1}{2}} = \1_{S_x} \:.
\end{align*}
The uniqueness of~$U_{xy}$ follows from Lemma~\ref{lemmapolar}.
Using~\eqref{Uxyex} in~\eqref{phiansatz} gives the result.
\QED

\subsection{Gaussian Wave Charts} \label{secgwc}
We now analyze whether the Gaussian charts constructed in Section~\ref{secgauss} also
give rise to wave charts. Our starting point is the local parametrization~$\Lambda$ in~\eqref{Lambda}.
Our strategy is to construct a mapping~$\psi \in \Lin(I, S_x) \oplus \Lin(J, S_x)$ such that~$M = R(\psi)$
(where~$R$ is again the mapping~\eqref{Rdef}). In other words, in block matrix notation,
we want to find~$\psi$ such that
\beq \label{Mcond}
M = \begin{pmatrix} \psi_I^\dagger \\[0.2em] \psi_J^\dagger \end{pmatrix} X \, \begin{pmatrix} \psi_I & \psi_J \end{pmatrix} \:.
\eeq
Considering the upper left block matrix entry of~$M$
and comparing with~\eqref{Lambda}, one finds that~$\psi_I$ must satisfy the equation
\beq \label{psiIcond}
\psi_I^\dagger \,X\, \psi_I = X + A \:.
\eeq
This equation can be solved with the spectral calculus: The first step is to set
\[ X+A = X\, \big(1+X^{-1} A \big) = X\, \sqrt{1+X^{-1} A}\: \sqrt{1+X^{-1} A} \:, \]
where the square root is defined as a power series in~$X^{-1}A$. Using that for any~$p \in \N$,
\[ X\, \big(X^{-1} A \big)^p = \big(A X^{-1} \big)^p\:X \:, \]
it follows that
\[ X\, \sqrt{1+X^{-1} A} = \sqrt{1+ A X^{-1}}\,X = \Big( \sqrt{1+X^{-1} A} \Big)^\dagger\, X\:. \]
We conclude that
\beq \label{factor}
X+A = \Big( \sqrt{1+X^{-1} A} \Big)^\dagger \,X\, \sqrt{1+X^{-1} A} \:,
\eeq
giving the explicit solution of~\eqref{psiIcond}
\beq \label{psiIex}
\psi_I = \sqrt{1+X^{-1} A} 
\:.
\eeq
Using~\eqref{factor} in~\eqref{Mform0}, we can read off that~$\psi_J$ is given explicitly by
\begin{align}
\psi_J &= \sqrt{1+X^{-1} A}\: (X+A)^{-1}\, B \notag \\
&= \sqrt{1+X^{-1} A}\; (1+X^{-1}\,A)^{-1}\,X^{-1}\, B \notag \\
&= \big(1+X^{-1} A \big)^{-\frac{1}{2}}\,X^{-1}\, B
\:. \label{psiJex}
\end{align}
With~\eqref{psiIex} and~\eqref{psiJex} we have found an explicit solution~$\psi=\psi_I+\psi_J$ of~\eqref{Mcond}.
Our findings are summarized as follows:
\begin{Prp} \label{prpgausspar}
In the Gaussian parametrization~$\Lambda$ in~\eqref{Lambda}, the spectral calculus gives rise to a canonical mapping
\begin{align}
\mathscr{W} \::\: &\big(\Symm(I) \oplus \Lin(I,J) \big) \cap B_\varepsilon(0) 
\rightarrow \Lin(I,S_x) \oplus \Lin(J, S_x) \notag \\
&(A,B) \mapsto \Big( \sqrt{1+X^{-1} A}, \: \big(1+X^{-1} A \big)^{-\frac{1}{2}}\,X^{-1}\, B \Big)
\label{ABform}
\end{align}
with the property that
\[ \Lambda(A,B) = R \big( \mathscr{W}(A,B) \big) \]
for all~$(A,B) \in \big(\Symm(I) \oplus \Lin(I,J) \big) \cap B_\varepsilon(0)$.
The mapping
\[ \phi := \mathscr{W} \circ \Lambda^{-1} \::\: \Omega \subset \F \rightarrow \Lin(I,S_x) \oplus \Lin(J, S_x) \]
with~$\Omega := \Lambda(B_\varepsilon(0))$ is a local chart of~$\F$.
\end{Prp} \noindent
The chart~$(\phi, \Omega)$ is referred to as the {\em{Gaussian wave chart}} about the point~$x \in M$.
Our construction is summarized by

\begin{tikzpicture}
\matrix (m) [matrix of math nodes,row sep=1.5cm,column sep=1.2cm,minimum width=2cm]
{
y \in \Omega \subset \F &	\Lambda^{-1}(y) \in \Symm(I) \oplus \Lin(I,J) & \phi(y) \in \Lin(\H, S_x)\:.  \\
	 };
\path[-stealth]
(m-1-1) edge node [above] {$\Lambda^{-1}$} (m-1-2)
(m-1-2) edge node [above] {$\mathscr{W}$} (m-1-3)
;
\end{tikzpicture}

\noindent
A more detailed analysis can be found in~\cite{kindermann}.

Let us analyze what this result means. Note that the operators~$A$ and~$B$ in
Proposition~\ref{prpgausspar} map subspaces of the Hilbert space~$\H$ into each other;
thus no indefinite inner product spaces appear. Nevertheless, the spin inner product
is important for understanding the formula in~\eqref{ABform}. The main observation
is the following simple lemma:
\begin{Lemma} \label{lemmasqrtsymm}
The operator~$\sqrt{1+X^{-1} A}$, regarded as an endomorphism of
the spin space~$S_x$, is symmetric,
\[ 
\sqrt{1+X^{-1} A} \in \Symm(S_x)\:. \]
\end{Lemma}
\Proof Using the formula~\eqref{Astar}, we obtain
\[ \big( X^{-1} A \big)^* = X^{-1} \big( X^{-1} A \big)^\dagger \,X = 
X^{-1} \big( A \,X^{-1} \big) \,X = X^{-1} A \:. \]
Hence all powers of~$X^{-1} A$ are also in~$\Symm(S_x)$.
Since the square root is defined by a power series, the result follows.
\QED
As a consequence of this lemma, the operator~${\mathscr{W}}$ maps to~$\Symm(S_x) \oplus \Lin(J, S_x)$.
Comparing with~\eqref{phichart}, we conclude that the Gaussian wave chart
satisfies the symmetry condition~\eqref{symmcond} used for the construction of the symmetric wave charts.
Using the uniqueness of the latter construction, we come to the following conclusion:
\begin{Prp} \label{prpcoincide}
The symmetric wave chart and the Gaussian wave chart about the point~$x \in M$ coincide.
\end{Prp}
This result gives a better understanding of the above constructions.
First of all, the fact that different constructions give the same wave charts shows
that our wave charts are canonical.
More technically, the symmetry condition which in~\eqref{symmcond} was introduced ad hoc,
gets a more convincing justification by Proposition~\ref{prpgausspar} and Lemma~\ref{lemmasqrtsymm},
where the condition~\eqref{symmcond} follows simply by rewriting the parametrization of
the Gaussian chart in terms of a wave chart.

\section{Example: Dirac Systems} \label{secex}
We now want to illustrate our results in concrete examples.
Knowing that symmetric wave charts and Gaussian wave charts coincide (see Proposition~\ref{prpcoincide}),
it suffices to consider the symmetric wave chart as computed in in Proposition~\ref{prpphi1}.
According to~\eqref{Psifix}, the corresponding gauge~$\Psi^\Omega_V$
(see Definition~\ref{defgauge}) is obtained by composing with a unitary operator~$U_x : S_x \rightarrow V$.
\subsection{Dirac Systems in Finite Spatial Volume}
We consider a system of non-interacting Dirac particles in finite spatial volume
(for basics on the Dirac equation we refer to~\cite{thaller} or standard textbooks
like~\cite{bjorken, schwabl2, peskin+schroeder}).
More precisely, let~$\scrM$ be the subset of Minkowski space
\beq \label{kasten}
\scrM := \R \times [-L,L]^3 \subset \R^{1,3}
\eeq
with periodic boundary conditions. The four-component Dirac spinors~$\psi(x)$ in Min\-kow\-ski space take values
in the spinor space, which we denote by~$S_x \scrM \simeq \C^4$.
The spinor space is endowed with an inner product~$\Sl .|. \Sr$ of signature~$(2,2)$
(which in physics is usually written as~$\Sl \psi|\phi \Sr = \overline{\psi} \phi$ with
the adjoint spinors~$\overline{\psi} = \psi^\dagger \gamma^0$), which we refer to as the
{\em{spin inner product}}.
For convenience, we extend the Dirac wave functions
to periodic functions in all of~$\R^{1,3}$, i.e.
\[ \psi \big(t, \vec{x} \big) = \psi \big(t, \vec{x}+\vec{v} \big)
\qquad \text{for all~$t \in \R, \vec{x} \in \R^3$
and~$\vec{v} \in (2L \Z)^3$}\:. \]
The scalar product on the Dirac solutions takes the usual form
\beq \label{printbox}
(\psi | \phi) := 2 \pi \int_{[-L,L]^3} \Sl \psi\big(t, \vec{x} \big) \:|\: \gamma^0\: \psi\big(t, \vec{x} \big) \Sr\: d^3x \:,
\eeq

Next, we make the plane-wave ansatz
\beq \label{planewave}
\psi_{\vec{k}as}(t,\vec{x}) = c\, e^{-i kx}\: \chi_{\vec{k}as} \qquad
\text{with~$\vec{k} \in \Big( \frac{\pi}{L}\: \Z \Big)^3$, $a\in \{1,2\}$ and~$s \in \{\pm 1\}$}\:,
\eeq
where~$c$ is a non-zero normalization constant
to be determined below. Here~$kx$ is the Minkowski inner product of the spacetime point~$x=(t,\vec{x})$
with four-momentum~$k$ on the mass shell,
\[ k := \big( s\: \omega(\vec{k}), \vec{k} \big) \qquad \text{with} \qquad
\omega(\vec{k}) := \sqrt{\big| \vec{k} \big|^2 + m^2} \:. \]
Moreover, the spinors~$\chi_{\vec{k}as}$ are solutions of the Dirac equation in momentum space
\beq \label{Dirack}
\big( \slashed{k}-m \big)\: \chi_{\vec{k}as} = 0 \:,
\eeq
which we choose to be pseudo-orthonormal with respect to the spin inner product, i.e.
\beq \label{pseudoONB}
\Sl \chi_{\vec{k}as} \,|\, \chi_{\vec{k} a' s} \Sr = s\: \delta_{a a'}\:.
\eeq
As a consequence, for fixed~$\vec{k}$ and~$s$, the integrand in~\eqref{printbox} is computed by
\begin{align*}
\Sl \chi_{\vec{k}as} \:|\: \gamma^0\: \chi_{\vec{k}a' s} \Sr
&= \frac{1}{2m} \Big( \Sl \slashed{k}\, \chi_{\vec{k}as} \:|\: \gamma^0\: \chi_{\vec{k}a' s} \Sr
+ \Sl \chi_{\vec{k}as} \:|\: \gamma^0\: \slashed{k}\, \chi_{\vec{k} a' s} \Sr \Big) \\
&= \frac{2 k^0}{2m} \;\Sl \chi_{\vec{k}as} \:|\: \chi_{\vec{k}a' s} \Sr 
= \frac{s\, \omega(\vec{k})}{m} \:s\: \delta_{a a'}= \frac{\omega(\vec{k})}{m} \: \delta_{a a'} \:.
\end{align*}
Moreover, this scalar product vanishes for fixed~$\vec{k}$ if the frequencies
of the waves have opposite signs,
\begin{align*}
\Sl \chi_{\vec{k}a+} \:|\: \gamma^0\: \chi_{\vec{k} a' -} \Sr
&= \frac{1}{2m} \Big( \Sl \big(\omega(\vec{k}) \,\gamma^0 -\vec{k} \vec{\gamma} \big)\, \chi_{\vec{k}a+} \:|\: \gamma^0\: \chi_{\vec{k}a' -} \Sr \\
&\qquad+ \Sl \chi_{\vec{k} a+} \:|\: \gamma^0\: \, \big(-\omega(\vec{k}) \,\gamma^0 -\vec{k} \vec{\gamma}  \big)\, \chi_{\vec{k} a' -} \Sr \Big) = 0 
\end{align*}
(where in the last line we used that~$[\gamma^0, \gamma^0] = 0 = \{\gamma^\alpha, \gamma^0\}$
for~$\alpha \in \{1,2,3\}$).
Using these formulas in~\eqref{printbox}, one concludes that the plane waves~\eqref{planewave}
are orthogonal. Moreover, the calculation
\[ (\psi_{\vec{k}as} | \psi_{\vec{k}as}) := 2 \pi \,|c|^2\, (2L)^3\:
\Sl \chi_{\vec{k} as} \,|\, \gamma^0\, \chi_{\vec{k} as} \Sr 
= 16\,\pi \,|c|^2\, L^3\: \frac{\omega(\vec{k})}{m} \]
shows that choosing the normalization constant as
\[ c = \sqrt{\frac{m}{\pi \omega(\vec{k})}}\: \frac{1}{4\,L^\frac{3}{2}}\:, \]
we obtain unit vectors. Our findings are summarized as follows:
\begin{Lemma} \label{lemmaplanewave}
In a three-dimensional box~\eqref{kasten} with periodic boundary conditions,
the Dirac wave functions
\beq \label{planewav2}
\psi_{\vec{k}as}(t,\vec{x}) = \sqrt{\frac{m}{\pi \omega(\vec{k})}}\: \frac{1}{4\,L^\frac{3}{2}}\:
e^{-i kx}\: \chi_{\vec{k}as}
\eeq
with~$\vec{k} \in (\pi \Z/L)^3$, $a\in \{1,2\}$ and~$s \in \{\pm 1\}$,
form an orthonormal basis of the Hilbert space of all Dirac solutions, endowed with the 
scalar product~\eqref{printbox}. Here~$\chi_{\vec{k}as}$ are pseudo-orthonormal solutions of the
Dirac equation in momentum space~\eqref{Dirack} and~\eqref{pseudoONB}.
\end{Lemma}

We now choose the Hilbert space~$\H$ as the subspace of the solution space
of all negative-energy solutions whose energy is above~$-1/\varepsilon$, i.e.\
\beq \label{Hplane}
\H := \text{span} \Big\{ \psi_{\vec{k}a-}(t,\vec{x}) \,\Big|\,
a \in \{1,2\} \text{ and } \omega \big(\vec{k} \big) < \frac{1}{\varepsilon} \Big\} \:.
\eeq
\begin{Lemma} The Hilbert space~$\H$ is finite-dimensional. Its dimension
has the following asymptotics for small~$\varepsilon$,
\[ f := \dim \H = \frac{8}{3 \pi^2}\: \bigg( \frac{L}{\varepsilon} \bigg)^3\;
\bigg( 1 + \O \Big(\frac{\varepsilon}{L} \Big) + \O \big(\varepsilon m \big) \bigg) \:. \]
\end{Lemma}
\Proof According to Lemma~\ref{lemmaplanewave}, two Dirac states of negative energy occupy a
volume of~$(\pi/L)^3$ in momentum space. As a consequence of the energy cutoff,
we must occupy a sphere of radius~$\sqrt{\varepsilon^{-2} - m^2}$ in momentum space.
Hence the number of states is counted by
\[ f = \frac{4 \pi}{3}\: \frac{1}{\varepsilon^3}\: 2\:\bigg(\frac{L}{\pi} \bigg)^3 
\bigg( 1 + \O \Big(\frac{\varepsilon}{L} \Big) + \O \big(\varepsilon m \big) \bigg) \:, \]
giving the result.
\QED
Having a finite-dimensional Hilbert space consisting of smooth wave functions, we can
define the local correlation operators without regularization operators, i.e.\
\[ F \::\: \scrM \rightarrow \F \:,\qquad (\psi \,|\, F(x)\, \phi) = \Sl \psi(x) \,|\, \phi(x) \Sr \quad
\forall\; \psi, \phi \in \H \:. \]
Finally, we define the universal measure as the push-forward of the Lebesgue measure on~$\scrM$,
\[ d\rho := F_*\big( \mu_\scrM \big) \qquad \text{with} \qquad d\mu_\scrM := d^4x \:. \]
We thus obtain a causal fermion system~$(\H, \F, \rho)$ of spin dimension~$n=2$.

\subsection{The Kernel of the Fermionic Projector in Finite Volume}
For the computations, it it is favorable to identify the
spin space~$S_x$ with the space~$S_x\scrM$ of Dirac spinors at the point~$x$ of Minkowski
space~$\scrM$. To this end, we introduce the evaluation operator~$e_x$ by
\beq \label{exdef}
e_x \::\: \H \rightarrow S_x\scrM\:, \qquad \psi \mapsto \psi(x)
\eeq
(here we use the fact that, according to~\eqref{Hplane}, the vectors of~$\H$ are not merely abstract vectors but linear
combinations of plane wave solutions of the Dirac equation, which can be evaluated at~$x \in \scrM$).
In~\cite[Section~1.2.4]{cfs} it is show that if~$e_x$ is surjective, then the spacetime point~$x$ is regular
(see~\cite[eq.~1.2.15]{cfs}).
Using this result, we now prove that our causal fermion system is regular  if the
dimension of the Hilbert space is sufficiently large:
\begin{Prp} \label{prpregular} If~$\dim \H \geq 4$, then the causal fermion system~$(\H, \F, \rho)$
is regular.
\end{Prp}
\Proof Assume that~$\dim \H \geq 4$. Then, since every momentum~$\vec{k}$ gives rise to two Dirac solutions,
at least two different momenta are occupied. According to~\eqref{Dirack} and~\eqref{pseudoONB},
for given~$\vec{k}$ the two spinors~$\chi_{\vec{k}a-}$ with~$a=1,2$ span the image of the matrix~$\slashed{k}+m$.
By direct computation, one sees that for two different momenta~$\vec{k}$ and~$\vec{k}'$, the span of the
images of the operators~$\slashed{k}+m$ and~$\slashed{k}'+m$ is four-dimensional. 
As a consequence, the corresponding four plane wave solutions~$\psi_{\vec{k}a-}(x)$ and~$\psi_{\vec{k}'a-}(x)$
evaluated at~$x$ are linearly independent. This implies that the evaluation operator~\eqref{exdef}
has rank four, giving the result.
\QED

From now on, we always assume that~$\dim \H \geq 4$, so that our causal fermion system
is regular. Restricting the evaluation operator to the subspace~$S_x \subset \H$, we obtain the mapping
\beq \label{identify}
e_x|_{S_x} \::\: S_x \rightarrow S_x\scrM \:.
\eeq
This mapping is indeed an isomorphism from the spin space to the spinor space
(for details see~\cite[Proposition~1.2.6]{cfs}),
making it possible to identify~$S_x$ and~$S_x\scrM$ as indefinite inner product spaces.
This identification is useful for bringing the objects of the causal fermion system
into a more explicit form, as we now explain in two examples.

\begin{Prp} \label{prpwave} Using the identification~\eqref{identify} of the spin spaces with the spinor spaces,
the wave evaluation operator~\eqref{Psixdef} coincides with the evaluation operator~\eqref{exdef},
\[ 
\Psi(x) \::\: \H \rightarrow S_x\scrM\:,\qquad u \mapsto e_x u = u(x) \:. \]
\end{Prp} \noindent
For the proof see~\cite[Proposition~1.2.6]{cfs},
choosing the regularization operator as the identity.

In the calculations it is most convenient to work with the kernel of the fermionic projector,
which for clarity we denote with indices~$\varepsilon$ and~$L$,
\beq \label{Pxydef}
P^{\varepsilon, L}(x,y) = \pi_x \,y|_{S_y} \::\: S_y \rightarrow S_x \:.
\eeq
\begin{Prp} Using the identification~\eqref{identify} of the spin spaces with the spinor spaces,
the kernel of the fermionic projector~\eqref{Pxydef} takes the form
\beq \label{Pxyplane}
P^{\varepsilon, L}(x,y) = \frac{1}{(2L)^3} \sum_{\stackrel{\vec{k} \in (\pi \Z/L)^3,}{\omega(\vec{k}) < \varepsilon^{-1}}}\;
\frac{1}{4 \pi \,\omega(\vec{k})}\: e^{-ik(x-y)}\;
(\slashed{k}+m) \bigg|_{k=\big(-\omega(\vec{k}), \vec{k} \big)}\:.
\eeq
\end{Prp}
\Proof According to~\cite[Proposition~1.2.7]{cfs}, under the identification~\eqref{identify}
the kernel of the fermionic projector takes the form
\beq \label{Pwave}
P^{\varepsilon, L}(x,y) = -\sum_{\vec{k}, a} |\psi_{\vec{k}a-}(x) \Sr \Sl \psi_{\vec{k}a-}(y)|
\eeq
(where we used a bra/ket notation and made use of the fact that in our example, there is no
regularization operator). Using the explicit form of the plane wave solutions~\eqref{planewav2},
we obtain
\[ P^{\varepsilon, L}(x,y) 
= -\frac{1}{(2L)^3} \sum_{\vec{k}, a} \frac{m}{2 \pi \,\omega(\vec{k})}\: e^{-ik(x-y)}\;
|\chi_{\vec{k}a-} \Sr \Sl \chi_{\vec{k}a-}| \:. \]
The bra/ket combination of the spinors~$\chi_{\vec{k}a-}$ can be calculated further.
Indeed, using that these spinors form a pseudo-orthonormal basis of the solution space
of the Dirac equation space (see~\eqref{Dirack} and~\eqref{pseudoONB}),
it is clear that the operator
\[ -\sum_{a=1}^2 |\chi_{\vec{k}a-} \Sr \Sl \chi_{\vec{k}a-}| \]
is an idempotent symmetric operator (with respect to the spin inner product)
whose image coincides with that of the operator~$\slashed{k}+m$. As a consequence,
\[ -\sum_{a=1}^2 |\chi_{\vec{k}a-} \Sr \Sl \chi_{\vec{k}a-}| = \frac{1}{2m}\: (\slashed{k}+m) \:. \]
This concludes the proof.
\QED

\subsection{Connection to the Kernels in Infinite Volume} \label{secinfvol}
In order to bring the kernel of the fermionic projector in~\eqref{Pxyplane} into a more
explicit form, it is useful to compare it with the corresponding kernel in infinite volume.
The unregularized kernel is the integral over the lower mass shell (see~\cite[Section~1.2.5]{cfs}),
\[ P(x,y) := \int \frac{d^4k}{(2 \pi)^4}\: (\slashed{k}+m)\: \delta(k^2-m^2)\: \Theta(-k^0)\:e^{-ik(x-y)} \:. \]
The simplest method for the regularization is to insert a momentum cutoff
(see also~\cite[Section~3.8.6~(B)]{cfs}),
\[ P^\varepsilon(x,y) := \int \frac{d^4k}{(2 \pi)^4}\: (\slashed{k}+m)\: \delta(k^2-m^2)\: \Theta(-k^0)\:
\Theta \big(1+\varepsilon k^0 \big) \:e^{-ik(x-y)} \:. \]

\begin{Prp} \label{prpkernel}
The unregularized kernel, the regularized kernel and the kernel in finite volume are related to each other by
\begin{align}
P^\varepsilon(x,y) &= \int_{-\infty}^\infty P\Big(x,\,y+\big(t,\vec{0} \big) \Big)\: 
\frac{1}{\pi t}\: \sin \Big( \frac{t}{\varepsilon} \Big)\: dt \label{conv1} \\
P^{\varepsilon,L}(x,y) &= \sum_{\vec{z} \in (2L \Z)^3} P^\varepsilon\Big(x,\,y+\big(0,\vec{z} \big) \Big) \:.
\label{conv2}
\end{align}
\end{Prp}
\Proof The momentum cutoff is realized by multiplying in momentum space with the
characteristic function~$\chi_{[-\varepsilon^{-1}, \varepsilon^{-1}]}(\omega)$ with~$\omega=k^0$.
Multiplication in momentum space corresponds to convolution in momentum space with the kernel
\[ \hat{\chi}_{[-\varepsilon^{-1}, \varepsilon^{-1}]}(t) := \int_{-\frac{1}{\varepsilon}}^{\frac{1}{\varepsilon}}
\frac{d\omega}{2 \pi} e^{-i \omega t} = \frac{i}{2 \pi t} \Big( e^{-\frac{it}{\varepsilon}} - 
e^{\frac{it}{\varepsilon}} \Big) = \frac{1}{\pi t}\: \sin \Big( \frac{t}{\varepsilon} \Big) \:. \]
This proves~\eqref{conv1}.

In order to derive~\eqref{conv2}, we rewrite~\eqref{Pxyplane} as
\begin{align*}
P^{\varepsilon, L}(x,y) 
&= \int_{-\infty}^\infty \frac{d\omega}{2 \pi}\; \frac{1}{(2L)^3} 
\sum_{\stackrel{\vec{k} \in (\pi \Z/L)^3,}{\omega(\vec{k}) < \varepsilon^{-1}}}\;
\delta(k^2-m^2)\; \Theta(-k^0)\: e^{-ik(x-y)}\; (\slashed{k}+m) \\
&= \int \frac{d^4k}{(2 \pi)^4}\;e^{-ik(x-y)}\;
\bigg( \Big( \frac{\pi}{L} \Big)^3 \sum_{\vec{q} \in (\pi \Z/L)^3} \delta^3 \big(\vec{k}-\vec{q} \big) \bigg) \\
&\qquad\qquad \times
\Big( (\slashed{k}+m)\:\delta(k^2-m^2)\; \Theta(-k^0)\: \Theta \big(1+\varepsilon k^0 \big) \Big) \:.
\end{align*}
Again using that multiplication in momentum space corresponds to convolution in position space,
one sees that~$P^{\varepsilon, L}$ is obtained from~$P^\varepsilon$ by convolution with the
spatial kernel
\begin{align*}
h(\vec{x}) &= \Big( \frac{\pi}{L} \Big)^3 \int \frac{d^3k}{(2 \pi)^3}
\sum_{\vec{q} \in (\pi \Z/L)^3} \delta^3 \big(\vec{k}-\vec{q} \big) \:e^{i \vec{k} \vec{x}} \\
&= \frac{1}{(2L)^3} \sum_{\vec{q} \in (\pi \Z/L)^3} e^{i \vec{q} \vec{x}}
= \sum_{\vec{z} \in (2L \Z)^3} \delta^3(\vec{x}-\vec{z}) \:,
\end{align*}
where in the last step we used the completeness relation for plane waves on the torus.
\QED
For clarity we remark that the sum in~\eqref{conv2} makes~$P^{\varepsilon,L}$
periodic in space with period~$2L$.

\subsection{Gauge Fixing of Wave Functions in Spacetime} \label{secfixst}
We now compute the gauge~\eqref{Psifix} for the symmetric wave chart~$\phi$ of Proposition~\ref{prpphi1}
more explicitly for our Dirac systems.
Although this gauge was derived under the assumption that~$\H$ is finite-dimensional,
all the formulas expressed in terms of the kernel of the fermionic projector
can be used in the infinite-dimensional setting of Section~\ref{secinfvol} just as well.
With this in mind, the following results apply
to the kernels with regularization in Proposition~\ref{prpkernel}, both in finite and infinite spatial volume.
For ease in notation, from now on we omit the indices~$\varepsilon$ and~$L$.

We begin with the {\em{massless case}}~$m=0$.
Then by symmetry it follows that
\beq \label{Pdiag}
P(x,x) = \alpha \gamma^0 \qquad \text{with} \qquad \alpha \in \R
\eeq
and thus
\[ U_{x,y} = \big( \gamma^0\, A_{xy}\, \gamma^0 \big)^{-\frac{1}{2}}\, \gamma^0\, P(x,y) \:. \]
Using that~$(\gamma^0)^2=\1$, we obtain
\[ \big( \gamma^0\, A_{xy}\, \gamma^0 \big)^p = \gamma^0\, A_{xy}^p\, \gamma^0 \]
for any~$p \in \N$. The spectral calculus yields that this relation holds also for any real~$p$.
Hence
\[ U_{x,y} = \big( \gamma^0\, A_{xy}^{-\frac{1}{2}}\, \gamma^0 \big)\, \gamma^0\, P(x,y)
= \gamma^0\, A_{xy}^{-\frac{1}{2}}\, P(x,y) \:. \]
Hence
\begin{align*}
\phi(y) = U_{x,y}\, \Psi(y) = \gamma^0\, A_{xy}^{-\frac{1}{2}}\, P(x,y) \: \Psi(y)\:.
\end{align*}
The resulting {\em{symmetric wave gauge}} is
\beq \label{disgauge}
\Psi^\Omega_V(y) = U_x \, \gamma^0\, A_{xy}^{-\frac{1}{2}}\, P(x,y) \: \Psi(y) \:.
\eeq

Before going on, we point out that the combination~$A_{xy}^{-\frac{1}{2}}\, P(x,y)$ is reminiscent of
the {\em{spin connection}} in~\cite{lqg}. Indeed, the spin connection
has the form (see~\cite[eq.~(3.42)]{lqg})
\beq \label{spinconnect}
D_{x,y} = e^{i \varphi_{xy}\, v_{xy}}\: A_{xy}^{-\frac{1}{2}}\: P(x,y) \:.
\eeq
where~$v_{xy}$ is the directional sign operator (see~\cite[Definition~3.15]{lqg}).
In simple terms, the factor~$e^{i \varphi_{xy}\, v_{xy}}$ introduces generalized $\text{SU}(2)$-phases
which are absent in~\eqref{disgauge}. These phases are important for the geometric
constructions in~\cite{lqg}. The drawback is that the spin connection~\eqref{spinconnect}
is {\em{not}} defined
for all spacetime points~$y$ in an open neighborhood of~$x$, but only for a more restrictive class of
spacetime points which satisfy the conditions subsumed in the notion of spin connectability
(see~\cite[Definition~3.17]{lqg}). With this in mind, the spin connection~\eqref{spinconnect}
cannot be used for constructing charts. The factor~$\gamma^0\, A_{xy}^{-\frac{1}{2}}\, P(x,y)$
in~\eqref{disgauge} can be understood as a simplified version of a spin connection,
which is insufficient for describing the geometry of spacetime, but which can nevertheless
be used for constructing distinguished gauges.

In the massless case, the kernel of the fermionic projector~$P(x,y)$ has only a vector
component (see~\eqref{Pxyplane} or the similar formulas in infinite volume).
Therefore, we can make the general ansatz
\beq \label{Puv}
P(x,y) = \slashed{u}(x,y) + i \slashed{\zeta}(x,y)
\eeq
with two Minkowski vectors~$u$ and~$\zeta$. In view of~\eqref{Pdiag},
\[ \slashed{u}(x,x) = \alpha\, \gamma^0 \qquad \text{and} \qquad \slashed{\zeta}(x,x) = 0 \:. \]
Moreover, we know that
\[ P(y,x) = P(x,y)^* = \slashed{u}(x,y) - i \slashed{\zeta}(x,y) \:. \]
Hence, omitting the arguments~$x$ and~$y$, we obtain for the closed chain
\beq \label{Axyform}
A_{xy} = u^2 + \zeta^2 - i\, [\slashed{u}, \slashed{\zeta}] \:.
\eeq
In the next lemma we compute the factor~$A^{-\frac{1}{2}}\, P(x,y)$ in~\eqref{disgauge}.
\begin{Lemma} \label{lemmamatrix} For the kernel of the fermionic given by~\eqref{Puv},
\begin{align*}
A_{xy}^{-\frac{1}{2}}\, P(x,y) &= \frac{1}{2} \bigg( 
\frac{\sqrt{\lambda_+} + \sqrt{\lambda_-}}{\sqrt{\lambda_+ \lambda_-}}
-\frac{\zeta^2 - i (u\zeta)}{\sqrt{u^2 \zeta^2 - 2 (u\zeta)^2}}
\: \frac{\sqrt{\lambda_+} - \sqrt{\lambda_-}}{\sqrt{\lambda_+ \lambda_-}} \bigg) \, \slashed{u} \\
& \quad\: + \frac{i}{2} \bigg( 
\frac{\sqrt{\lambda_+} + \sqrt{\lambda_-}}{\sqrt{\lambda_+ \lambda_-}}
-\frac{u^2 - i (u\zeta)}{\sqrt{u^2 \zeta^2 - 2 (u\zeta)^2}}
\: \frac{\sqrt{\lambda_+} - \sqrt{\lambda_-}}{\sqrt{\lambda_+ \lambda_-}} \bigg) \, \slashed{\zeta}\:,
\end{align*}
where
\beq \label{lpm}
\lambda_\pm = u^2 + \zeta^2 \pm 2 \sqrt{u^2\, \zeta^2 - (u\zeta)^2} \:.
\eeq
\end{Lemma}
\Proof The calculation
\begin{align*}
&\big( A_{xy} - u^2-\zeta^2 \big)^2 = - [\slashed{u}, \slashed{\zeta}]^2
= -\slashed{u} \slashed{\zeta} \slashed{u} \slashed{\zeta} - \slashed{\zeta} \slashed{u} \slashed{\zeta} \slashed{u}
+ 2 u^2\, \zeta^2 \\
&=-\big(2\,(u\zeta)\, \slashed{u} \slashed{\zeta} - u^2 \zeta^2 \big)
- \big(2\,(u\zeta)\, \slashed{\zeta} \slashed{u} - u^2 \zeta^2 \big) + 2 u^2\, \zeta^2
= -4\, (u \zeta)^2 + 4 u^2\, \zeta^2
\end{align*}
shows that the matrix~$A_{xy}$ has the eigenvalues~$\lambda_\pm$ as given in~\eqref{lpm}.
The corresponding spectral projection operators are given by
\[ E_\pm = \frac{1}{2} \Big( \1 \pm \frac{A_{xy} - \lambda_\mp}{\lambda_+-\lambda_-} \Big) 
= \frac{1}{2} \Big( \1 \mp \frac{i\, [\slashed{u}, \slashed{\zeta}]}{2 \sqrt{u^2 \zeta^2 - (u \zeta)^2}} \Big) \:, \]
where in the last step we used~\eqref{Axyform} and~\eqref{lpm}.
The spectral calculus gives
\[ A_{xy}^{-\frac{1}{2}} \, P(x,y) = \sum_{s=\pm} \lambda_s^{-\frac{1}{2}}\, E_s\: P(x,y) \:. \]
Substituting~\eqref{Puv} and applying the relations
\begin{align*}
[\slashed{u}, \slashed{\zeta}]\, \slashed{u} &= \slashed{u} \slashed{\zeta} \slashed{u}
- u^2\, \slashed{\zeta} 
= \big(2\,(u\zeta)\, \slashed{u} - u^2\, \slashed{\zeta} \big) - u^2\, \slashed{\zeta} \\
&= 2\,(u\zeta)\, \slashed{u} - 2 u^2\, \slashed{\zeta} \\
[\slashed{u}, \slashed{\zeta}]\, \slashed{\zeta} &=
-2\,(u\zeta)\, \slashed{\zeta} + 2 \zeta^2\, \slashed{u}
\end{align*}
gives the result.
\QED

We next analyze this result in an expansion near the diagonal~$x=y$. To this end, we make the ansatz
\beq \label{uvex}
u = \alpha \gamma^0 + \tau\, u_1 + \O\big(\tau^2 \big) \:,\qquad
\zeta = \tau\, \zeta_1 + \O\big(\tau^2 \big)
\eeq
with a real expansion parameter~$\tau$.
A straightforward computation (which we carried out with the help of Mathematica) gives the
following result:
\begin{Prp} \label{prpgP} For~$P(x,y)$ as in~\eqref{Puv} with~$u$ and~$\zeta$ according to~\eqref{uvex},
\beq \label{gP}
\gamma^0 \, A_{xy}^{-\frac{1}{2}}\, P(x,y)
= \1 - \tau\, \gamma^0\, \big( \vec{u} \vec{\gamma} \big) + \tau\: \frac{i \zeta_1^0}{|\alpha|}\: \1
+ \O\big(\tau^2 \big) \:.
\eeq
\end{Prp}
Let us explain the above results. We begin with Proposition~\ref{prpgP}.
Writing the matrix in~\eqref{gP} in the form~$\1 + \tau A + \O(\tau^2)$,
the fact that~$A$ is antisymmetric (with respect to the spin inner product) shows that this matrix is unitary.
Next, one sees that only the spatial component of~$u$ and only the time component of~$\zeta$ enter~\eqref{gP}.
More precisely, the time component of~$\zeta$ gives a phase factor, whereas the spatial component of~$u$
gives a bilinear contribution. These contributions clearly depend on the regularization scale~$\varepsilon$.
The expansion in Proposition~\ref{prpgP} is justified only if the difference vector~$y-x$ is as small as the
regularization scale. On larger scales, one must work instead with the formulas of Lemma~\ref{lemmamatrix}.
In general terms, the matrix in~\eqref{gP} is a unitary mapping from the spinor spaces at~$y$ to~$x$,
which depends on the difference vector~$y-x$ and on the regularization.

In the resulting symmetric wave gauge~$\Psi^\Omega_V$ in~\eqref{disgauge}, this matrix is multiplied by~$\Psi(y)$,
which is composed of the plane-wave solutions of the Dirac equation at the spacetime point~$y$
(see Proposition~\ref{prpwave}). The point of interest is that {\em{gauge phases drop out}}
of~$\Psi^\Omega_V$. This can be seen explicitly from the transformation law under
gauge transformations~\eqref{gauge1} and~\eqref{gauge2}, which implies that
\begin{align*}
\Psi(y) &\rightarrow e^{i \Lambda(y)}\: \Psi(y) \\
P(x,y) &\rightarrow e^{i \Lambda(x) - i \Lambda(y)}\: P(x,y) \\
A_{xy} &\rightarrow A_{xy} \\
\Psi^\Omega_V(y) &\rightarrow \Psi^\Omega_V(y)\:. 
\end{align*}
Thus the local gauge freedom of electrodynamics is completely fixed.

Due to the phases depending on~$y-x$ in~\eqref{gP}, the symmetric wave gauge cannot be identified
with any of the usual gauges of electrodynamics (like the Lorenz, Coulomb 
or general $R_xi$ gauges).
Instead, the local phases are determined by the detailed form of the regularization.

We finally explain how the above findings generalize to the {\em{massive case}}
$m>0$. In this case, the regularized kernel~$P(x,y)$ also involves a scalar component,
making all the formulas more complicated. However, for~$y-x$ on the Planck scale,
the scalar component is smaller than the vector component by a scaling factor of~$\varepsilon m$.
Therefore, the result of Proposition~\ref{prpgP} is still valid, up to small correction terms.
With this in mind, all our qualitative results remain valid, but the detailed form of the
gauge fixing is more involved.

\subsection{Gauge Fixing of the Perturbation Expansion} \label{secfixpe}
We now consider the situation that the Dirac wave functions are perturbed by an external electromagnetic
potential~$A$. 
Before beginning, we briefly explain how the electromagnetic potential comes into play
in the analysis of the dynamics of causal fermion systems.
As explained in Section~\ref{secquantumcfs}, it is a central idea 
behind causal fermion systems to describe the physical system
purely in terms of the ensemble of wave functions.
Implementing this idea in a gauge-invariant way leads to the definition
of causal fermion systems (see Definition~\ref{defparticle}).
The dynamics of a causal fermion systems is described by a variational
principle for the measure~$\rho$, referred to as the {\em{causal action principle}}
(see for example~\cite[Section~1.1]{cfs}). This action principle can be understood
as describing an interaction of all the physical wave functions of the system.
In order to write this interaction in a more tractable form, it is very helpful to
describe the collective behavior of all the physical wave functions by
bosonic potentials. This procedure has been carried out systematically in~\cite{cfs},
leading to the so-called {\em{continuum limit}} analysis where the interaction is described
effectively by classical bosonic gauge fields coupled to fermionic wave functions.
In the present paper, we do not enter the analysis of the causal action principle.
Instead, we simply perturb the system of Dirac wave functions by an external
electromagnetic potential~$A$ and analyze how the resulting causal fermion system changes.

It is most convenient to begin with the perturbation of the wave evaluation operator~$\Psi$.
Always denoting the perturbed objects by a tilde, to first order we obtain
\beq \label{firstorder}
\tilde{\Psi}(x) = \Psi(x) - \big( s_m \slashed{A} \Psi \big)(x)
= \Psi(x) - \int s_m(x,y) \slashed{A}(y)\,\Psi(y)\: d^4y \:,
\eeq
where~$s_m$ is a Dirac Green's operator. To higher order, one has similar formulas involving
several Green's operators (for a systematic treatment see for example~\cite{norm}).
Here we do not need to enter the details of the perturbation expansion.
It suffices to note that the perturbation expansion respects the gauge symmetry in the sense
that a pure gauge potential~$\slashed{A}(x) = \Pdd \Lambda(x)$ gives rise to a local phase
transformation,
\beq \label{Psiphase}
\tilde{\Psi}(x) = e^{i \Lambda(x)}\: \Psi(x) \:.
\eeq
To first order, this can be verified directly from~\eqref{firstorder} using the computation
\begin{align*}
\tilde{\Psi}(x) &= \Psi(x) - \big( s_m (\Pdd \Lambda) \Psi \big)(x)
= \Psi(x) +i \big( s_m [i \Pdd- m,  \Lambda] \Psi \big)(x) \\
&= \Psi(x) +i \Lambda(x)\: \Psi(x) = e^{i \Lambda(x)}\: \Psi(x) + \O\big( \Lambda^2 \big) \:.
\end{align*}

Once we know~$\tilde{\Psi}$, all the other relevant objects can
computed in a straightforward way.
In particular, the perturbed local correlation operator and the kernel of the fermionic
projector are given by (for details see~\cite[Lemma~1.1.3]{cfs})
\[ \tilde{F}(x) = - \tilde{\Psi}(x)^* \tilde{\Psi}(x) \qquad \text{and} \qquad
\tilde{P}(x,y) = -\tilde{\Psi}(x) \tilde{\Psi}(y)^* \:. \]
We can also perturb only one of the factors in the kernel of the fermionic projector. We use the notation
\[ P \big(x, \tilde{F}(y) \big) := -\Psi(x) \tilde{\Psi}(y)^* \:. \]

In order to fix the gauge in the perturbation expansion, 
one should note that~$\tilde{F}(x)$ is again an operator in~$\F$. Therefore,
we can work again with~$\phi$ in Proposition~\ref{prpphi1}
choosing~$y=\tilde{F}(x)$, i.e.\
\[ \tilde{\phi}(x) := \big( P(x,x)^{-1}\, A_{x\, \tilde{F}(x)}\, P(x,x)^{-1} \big)^{-\frac{1}{2}}\, P(x,x)^{-1}\,
P\big(x,\tilde{F}(x) \big) \: \tilde{\Psi}(x) \:. \]
Using again that in our Dirac examples, $P(x,x)$ has the form~\eqref{Pdiag}, we can simplify
this formula according to~\eqref{disgauge} to obtain the perturbation expansion
in the {\em{symmetric wave gauge}}
\beq \label{disgauge2}
\tilde{\Psi}^\Omega_V(x) = U_x \, \gamma^0\, A_{x\, \tilde{F}(x)}^{-\frac{1}{2}}\, P\big(x,\tilde{F}(x) \big) \:
\tilde{\Psi}(x) \:.
\eeq

In order to understand what this formula means, it is useful to choose an orthonormal
basis~$u_1, \ldots, u_4$ of the subspace~$S_x \subset \H$ (orthonormal with respect to the
scalar product~$\la .|. \ra_\H$). Then for any~$y \in \F$,
\beq \label{Pyxloc}
P(y,x) = \pi_y x|_{S_x} = \sum_{a=1}^4 \pi_y u_a \, \la u_a | x \big|_{S_x} = -\sum_{a=1}^4 |u_a(y) \Sr \Sl u_a(x)| \:,
\eeq
where in the last step as in~\eqref{Pwave} we again applied~\cite[Proposition~1.2.7]{cfs}
and used the identification~\eqref{identify}. Choosing~$y=\tilde{F}(x)$, we obtain the simple formulas
\begin{align}
P\big(x,\tilde{F}(x) \big) &= -\sum_{a=1}^4 |u_a(x) \Sr \Sl \tilde{u}_a(x)| \label{PxF} \\
A_{x,\tilde{F}(x)} &= -\sum_{a,b=1}^4 
|u_a(x) \Sr \Sl \tilde{u}_a(x)| \tilde{u}_b(x) \Sr \Sl u_a(x) | \:. \label{AxF}
\end{align}
This shows that the formula~\eqref{disgauge2} can be expressed purely in terms of
the unperturbed and perturbed wave functions~$u_a$ and~$\tilde{u}_a$, all evaluated at the spacetime point~$x$.

This raises the question how the wave functions~$u_a$ and~$\tilde{u}_a$ look like.
Indeed, this can be read off from~\eqref{Pyxloc}:
\begin{Lemma} For Dirac systems in Minkowski space,
\begin{align}
u_a(y) &= P(y,x)\: \chi_a \label{uay} \\
\tilde{u}_a(x) &= P\big( \tilde{F}(x),\, x \big)\: \chi_a \:, \label{uat}
\end{align}
where~$\chi_a \in S_x\scrM$ are the spinors
\beq \label{chiadef}
\chi_a = \frac{1}{\alpha}\: \gamma^0\, u_a(x) \:,\qquad a=1,\ldots, 4\:.
\eeq
\end{Lemma}
\Proof Multiplying~\eqref{Pyxloc} by a spinor~$\chi \in S_x \scrM$, we obtain
\beq \label{lincomb}
\sum_{b=1}^4 c_b\: u_b(y) = P(y,x) \,\chi
\eeq
with coefficients~$c_b = -\Sl u_b(x)| \chi \Sr$.
Hence the Dirac wave functions~$u_a(y)$ are obtained by multiplying~$P(y,x)$ with suitable spinors.
In order to prove~\eqref{uat}, it remains to verify that choosing~$\chi=\chi_a$ according to~\eqref{chiadef},
the linear combination on the left of~\eqref{lincomb} gives the wave function~$u_a$.
To this end, it suffices to evaluate~\eqref{lincomb} for~$y=x$,
\[ \sum_{b=1}^4 c_b\: u_b(x) = P(x,x) \, \chi = \alpha\, \gamma^0 \, \chi\:, \]
where in the last step we applied~\eqref{Pdiag}. Using~\eqref{chiadef} gives
\[ \sum_{b=1}^4 c_b\: u_b(x) = u_a(x) \:, \]
concluding the proof of~\eqref{uay}.

The identity~\eqref{uat} follows from~\eqref{uay} by comparing~\eqref{Pyxloc} with~\eqref{PxF}.
\QED

We now compute the wave functions corresponding to the vectors~$u_a$ in the
symmetric wave gauge.
\begin{Prp} In the gauge~\eqref{disgauge2}, the vectors~$u_1, \ldots, u_4$ which form an
orthonormal basis of~$S_x$ have the form
\beq \label{gaugeua}
\tilde{\Psi}^\Omega_V(x) \, u_a = U_x \, \gamma^0\, A_{x\, \tilde{F}(x)}^{\frac{1}{2}}\, \chi_a \:.
\eeq
\end{Prp}
\Proof Using~\eqref{uat} in~\eqref{disgauge2} gives
\begin{align*}
\tilde{\Psi}^\Omega_V(x) \, u_a
&= U_x \, \gamma^0\, A_{x\, \tilde{F}(x)}^{-\frac{1}{2}}\, P\big(x,\tilde{F}(x) \big) \:\tilde{u}_a(x) \\
&= U_x \, \gamma^0\, A_{x\, \tilde{F}(x)}^{-\frac{1}{2}}\, P\big(x,\tilde{F}(x) \big) \:
P\big( \tilde{F}(x),\, x \big)\: \chi_a \\
&= U_x \, \gamma^0\, A_{x\, \tilde{F}(x)}^{-\frac{1}{2}}\, A_{x\, \tilde{F}(x)}\, \chi_a \:,
\end{align*}
giving the result.
\QED

Our gauge-fixing procedure can be understood directly by comparing the
wave functions of the vectors~$u_a$ without gauge fixing~\eqref{uat}
with those in the symmetric wave gauge~\eqref{gaugeua}.
In~\eqref{uat}, the wave functions are modified by the electromagnetic
potential. In particular, for a gauge transformation, this gives rise to the
local phase in~\eqref{Psiphase}. The formula~\eqref{gaugeua}, on the other hand,
involves instead of~$P(\tilde{F}(x),x)$ the matrix~$A_{x,\tilde{F}(x)}^{\frac{1}{2}}$.
This matrix does {\em{not}} involve gauge phases,
because the close chain is gauge invariant according to~\eqref{AxF}.
The matrix~$\gamma^0$ is needed in order to get agreement of~\eqref{uat} and~\eqref{gaugeua}
in the case when no electromagnetic potential is present.

In this way, our gauge fixing procedure brings the wave functions corresponding to the vectors~$u_a$
into a canonical form. The point is that by doing so, the $\U(2,2)$-gauge freedom at
the spacetime point~$x$ is exhausted completely.
Therefore, the wave functions corresponding to all other vectors in~$\H$ at~$x$
are also determined uniquely.

We conclude by giving an intuitive picture of how the Dirac waves~$u_a$ look like in position space
and outline the methods for analyzing their perturbations.
According to~\eqref{uay}, the spacetime dependence of these waves is the same as that of
the kernel of the fermionic projector~$P(y,x)$ for fixed~$x$.
The unregularized kernel~$P(y,x)$ has {\em{singularities}} if~$y$ lies {\em{on the light cone}} centered at~$x$
(for details see for example~\cite[Section~1.2]{cfs}).
Due to the regularization, these singularities are mollified on the scale~$\varepsilon$.
This means that, for small~$\varepsilon$, the Dirac waves~$u_a(y)$ are peaked near the
light cone centered at~$y$. Qualitatively speaking, these waves can be regarded as
wave packets of negative frequency which are as far as possible localized at time~$t=x^0$
at the spatial point~$\vec{x}$.
Clearly, in view of Hegerfeldt's theorem~\cite{hegerfeldt}
(see also~\cite[Section~1.8.3]{thaller}), wave packets of negative frequency cannot
be localized in space. This is also apparent here because, similar to the Feynman propagator,
the distribution~$P(y,x)$ does have a contribution if~$x$ and~$y$ are spatially separated,
but this contribution decays exponentially in the spatial distance.
More details on the waves~$u_a$ and related results on Dirac systems in Minkowski space
can be found in~\cite{oppio}.

The {\em{light-cone expansion}} is a powerful computational tool for analyzing
the kernel~$\tilde{P}(x,y)$ in position space (see~\cite{firstorder, light} or
the introduction in~\cite[Section~2.2]{cfs}). The resulting formulas show that the electromagnetic
potential changes~$P(x,y)$ by gauge phases and also by contributions
involving the field tensor and its derivatives. More precisely, the unregularized distribution~$\tilde{P}(x,y)$
can be expressed by an infinite sum of distributions which have singularities on the light cone,
each multiplied by an integral over potentials or fields along the line segment~$\overline{xy}$.
The regularized kernel is then obtained by mollification (for details see~\cite[Appendix~F]{cfs}).
The light-cone expansion of~$P(x, \tilde{F}(y))$ is more involved because it typically involves
{\em{unbounded}} line integral along the straight line joining the points~$x$ and~$y$.
This is worked out in~\cite[Appendix~F]{pfp}; see also~\cite[Lemma~5.1]{action}.
However, these results give information on~$P(x, \tilde{F}(x))$ only if~$x \neq y$.
Therefore, these results unfortunately do not apply to the regularized kernel~$P(x, \tilde{F}(x))$
as needed for the symmetric wave gauge~\eqref{disgauge2}.
At present, the only rigorous result is that a pure gauge potential~$\slashed{A}=\Pdd \Lambda$
gives rise to a gauge phase,
\[ P(x, \tilde{F}(x)) = e^{-i \Lambda(x)}\: P(x,x) \:. \]
This suggests that the leading order in~$\varepsilon/l_{\text{\tiny{macro}}}$
(where~$l_{\text{\tiny{macro}}}$ denotes the macroscopic length scale determined by the
Compton scale and typical wave lengths of the electromagnetic field) should also
simply give a gauge phase. It seems a promising strategy for computing the 
higher orders in an expansion~$\varepsilon/l_{\text{\tiny{macro}}}$
to work in momentum space (similar to~\cite[Section~3]{firstorder}) and to integrate
over both the incoming and outgoing momenta.
However, the detailed computations are somewhat technical and go beyond the scope of the present paper.

\Thanks{{{\em{Acknowledgments:}} We would like to thank Marco Oppio and the referees for helpful comments on the
manuscript. We are grateful to Tilo Wettig for co-advising the master thesis~\cite{kindermann}.


\begin{thebibliography}{10}

\bibitem{cfsweblink}
\emph{Link to web platform on causal fermion systems:
  www.causal-fermion-system.com}.

\bibitem{bjorken}
J.D. Bjorken and S.D. Drell, \emph{Relativistic {Q}uantum {M}echanics},
  McGraw-Hill Book Co., New York, 1964.

\bibitem{gauge}
F.~Finster, \emph{Derivation of local gauge freedom from a measurement principle},
  arXiv:funct-an/9701002, Photon and Poincare Group (V.~Dvoeglazov, ed.), Nova
  Science Publishers, 1999, pp.~315--325.

\bibitem{u22}
\bysame, \emph{Local {$\rm U(2,2)$} symmetry in relativistic quantum
  mechanics}, arXiv:hep-th/9703083, J. Math. Phys. \textbf{39} (1998), no.~12,
  6276--6290.

\bibitem{firstorder}
\bysame, \emph{Light-cone expansion of the {D}irac sea to first order in the
  external potential}, arXiv:hep-th/9707128, Michigan Math. J. \textbf{46}
  (1999), no.~2, 377--408.

\bibitem{light}
\bysame, \emph{Light-cone expansion of the {D}irac sea in the presence of
  chiral and scalar potentials}, arXiv:hep-th/9809019, J. Math. Phys.
  \textbf{41} (2000), no.~10, 6689--6746.

\bibitem{pfp}
\bysame, \emph{The {P}rinciple of the {F}ermionic {P}rojector}, hep-th/0001048,
  hep-th/0202059, hep-th/0210121, AMS/IP Studies in Advanced Mathematics,
  vol.~35, American Mathematical Society, Providence, RI, 2006.

\bibitem{continuum}
\bysame, \emph{Causal variational principles on measure spaces},
  arXiv:0811.2666 [math-ph], J. Reine Angew. Math. \textbf{646} (2010),
  141--194.

\bibitem{cfs}
\bysame, \emph{The {C}ontinuum {L}imit of {C}ausal {F}ermion {S}ystems},
  arXiv:1605.04742 [math-ph], Fundamental Theories of Physics, vol. 186,
  Springer, 2016.

\bibitem{action}
\bysame, \emph{The causal action in {M}inkowski space and surface layer
  integrals}, arXiv:1711.07058 [math-ph] (2017).

\bibitem{nrstg}
\bysame, \emph{Causal fermion systems: A primer for {L}orentzian geometers},
  arXiv:1709.04781 [math-ph], J. Phys.: Conf. Ser. \textbf{968} (2018), 012004.

\bibitem{perturb}
\bysame, \emph{Perturbation theory for critical points of causal variational
  principles}, arXiv:1703.05059 [math-ph], to appear in Adv. Theor. Math. Phys.
  (2020).

\bibitem{lqg}
F.~Finster and A.~Grotz, \emph{A {L}orentzian quantum geometry},
  arXiv:1107.2026 [math-ph], Adv. Theor. Math. Phys. \textbf{16} (2012), no.~4,
  1197--1290.

\bibitem{review}
F.~Finster and M.~Jokel, \emph{Causal fermion systems: An elementary
  introduction to physical ideas and mathematical concepts}, arXiv:1908.08451
  [math-ph], {P}rogress and {V}isions in {Q}uantum {T}heory in {V}iew of
  {G}ravity (F.~Finster, D.~Giulini, J.~Kleiner, and J.~Tolksdorf, eds.),
  Birkh\"auser Verlag, Basel, 2020, pp.~63--92.

\bibitem{topology}
F.~Finster and N.~Kamran, \emph{Spinors on singular spaces and the topology of
  causal fermion systems}, arXiv:1403.7885 [math-ph], Mem. Amer. Math. Soc.
  \textbf{259} (2019), no.~1251, v+83 pp.

\bibitem{dice2014}
F.~Finster and J.~Kleiner, \emph{Causal fermion systems as a candidate for a
  unified physical theory}, arXiv:1502.03587 [math-ph], J. Phys.: Conf. Ser.
  \textbf{626} (2015), 012020.

\bibitem{intro}
F.~Finster, J.~Kleiner, and J.-H. Treude, \emph{An {I}ntroduction to the
  {F}ermionic {P}rojector and {C}ausal {F}ermion {S}ystems}, in preparation,
  https://causal-fermion-system.com/intro-public.pdf.

\bibitem{banach}
F.~Finster and M.~Lottner, \emph{Banach manifold structure and jet spaces in
  infinite-dimensional causal fermion systems}, in preparation.

\bibitem{norm}
F.~Finster and J.~Tolksdorf, \emph{Perturbative description of the fermionic
  projector: Normalization, causality and {F}urry's theorem}, arXiv:1401.4353
  [math-ph], J. Math. Phys. \textbf{55} (2014), no.~5, 052301.

\bibitem{hegerfeldt}
G.C. Hegerfeldt, \emph{Remark on causality and particle localization}, Phys.
  Rev. D \textbf{10} (1974), 3320--3321.

\bibitem{kindermann}
S.~Kindermann, \emph{{Geometrie und Eichfixierung des kausalen Fermionsystems
  in endlichem Volumen}}, Masterarbeit Physik, Universit\"at Regensburg (2019).

\bibitem{landau3}
L.D. Landau and E.M. Lifshitz, \emph{Quantum {M}echanics: {N}on-{R}elativistic
  {T}heory.}, {C}ourse of {T}heoretical {P}hysics, {V}ol. 3. {T}ranslated from
  the {R}ussian by {J.B. Sykes and J.S. Bell}, Addison-Wesley Series in
  Advanced Physics, Pergamon Press Ltd., London-Paris, 1958.

\bibitem{landau2}
\bysame, \emph{The {C}lassical {T}heory of {F}ields}, Revised second edition.
  Course of Theoretical Physics, Vol. 2. Translated from the Russian by M.
  Hamermesh, Pergamon Press, Oxford, 1962.

\bibitem{oppio}
M.~Oppio, \emph{On the mathematical foundations of causal fermion systems in
  {M}inkowski spacetime}, arXiv:1909.09229 [math-ph] (2019).

\bibitem{peskin+schroeder}
M.E. Peskin and D.V. Schroeder, \emph{An {I}ntroduction to {Q}uantum {F}ield
  {T}heory}, Addison-Wesley Publishing Company Advanced Book Program, Reading,
  MA, 1995.

\bibitem{sakurai}
J.J. Sakurai and J.~Napolitano, \emph{Advanced {Q}uantum {M}echanics}, second
  ed., Addison-Wesley Publishing Company, 1994.

\bibitem{schwabl2}
F.~Schwabl, \emph{Advanced {Q}uantum {M}echanics}, third ed., Springer-Verlag,
  Berlin, 2005.

\bibitem{thaller}
B.~Thaller, \emph{The {D}irac {E}quation}, Texts and Monographs in Physics,
  Springer-Verlag, Berlin, 1992.

\end{thebibliography}
\providecommand{\bysame}{\leavevmode\hbox to3em{\hrulefill}\thinspace}
\providecommand{\MR}{\relax\ifhmode\unskip\space\fi MR }
\providecommand{\MRhref}[2]{%
  \href{http://www.ams.org/mathscinet-getitem?mr=#1}{#2}
}
\providecommand{\href}[2]{#2}


\end{document}